\documentclass{ar-1col-S2O}
\usepackage[numbers,compress]{natbib}
\usepackage{url}
\usepackage{amsmath}
\usepackage{caption}
\usepackage{subcaption}
\usepackage{multirow}
\usepackage{makecell}
\usepackage{rotating}

\jname{Submitted to Annu. Rev. Nucl. Part. Sci. }
\jvol{XX}
\jyear{2024}
\doi{10.1146/annurev-nucl-102622-013547}

\usepackage{defs}

\begin{document}

\markboth{B. Clerbaux and C. Gwilliam}{Searches for New High-Mass Fermionic Resonances at the LHC}

\title{Searches for New High-Mass Resonances Decaying to Fermions at the LHC}

\author{Barbara Clerbaux$^1$, Carl Gwilliam$^2$
\affil{$^1$Interuniversity Institute for High Energies (IIHE), Universi\'{e} libre de Bruxelles, ULB, Brussels, Belgium; email: barbara.clerbaux@ulb.be}
\affil{$^2$Department of Physics, University of Liverpool, Liverpool, UK; email: C.Gwilliam@liverpool.ac.uk}}

\begin{abstract}

Accelerator searches for new resonances have a long-standing history of discoveries that have driven advances in our understanding of nature.  Since 2010, the Large Hadron Collider (LHC) has probed previously inaccessible energy scales, enabling searches for new heavy resonances predicted by a wide range of theories beyond the standard model (BSM).  In particular, resonance decays into fermionic final states are often seen as golden channels since they provide a clear signal,
typically a peak in the invariant mass of the decay products over a smoothly-falling background distribution.  This review summarises the key concepts of the experimental searches for new resonances decaying to fermions, in the context of the BSM theories that motivate them, and presents the latest results of the ATLAS and CMS experiments, focusing on the complete LHC run-2 dataset.  Future prospects at the high-luminosity LHC and potential future colliders are also surveyed.
\end{abstract}

\maketitle

\tableofcontents 

\section{INTRODUCTION}



The standard model (SM) of particle physics has been immensely successful in describing the
fundamental building blocks of matter and the interactions between them across many orders of magnitude in energy.  With the momentous discovery of the Higgs boson in 2012~\cite{ATLAS:2012yve,CMS:2012qbp} the final piece of the model was put in place.  However, despite this theoretical and experimental triumph, many questions remain unanswered.   There is no explanation for the family structure of the fermions with their huge range of masses, while the Higgs boson is suspiciously light. 
On the cosmological scale, it is unable to explain the baryon asymmetry of the universe nor provide a particle origin for the mysterious dark matter (DM) that dominates it. 

Given this, it is clear that the SM cannot be the complete picture of nature and must instead be viewed as a low-energy approximation, albeit an incredibly predictive one, of some more fundamental theory.  Fortunately, many theoretically compelling extensions of the SM, with varying degrees of complexity, have been proposed to address one or more of these shortcomings.  The vast majority of these beyond-the-SM (BSM) theories predict new heavy particles, ranging from new fundamental building blocks of fermionic matter, via additional Higgs or gauge bosons, to more exotic entities such as leptoquarks.    

Searches for new resonances produced at particle accelerators have historically heralded major advances in our understanding of nature.  The detection of the $J/\psi$~\cite{SLAC-SP-017:1974ind,E598:1974sol} in 1974 proved the existence of the predicted charm quark, heralding the so-called ``November'' revolution~\cite{Khare:1999zz}.  This was followed shortly afterwards by the unexpected discoveries of the tau lepton~\cite{Perl:1975bf} in 1975 and the $\Upsilon$~\cite{E288:1977xhf} in 1977, indicating the existence of a third generation of matter particles. In the gauge sector, the observation of the $W^{\pm}$ and $Z$ bosons~\cite{UA1:1983crd,UA2:1983tsx,UA1:1983mne,UA2:1983mlz} in 1983 confirmed the electroweak (EW) unification. It is worth noting that all these particles were discovered through their decays to fermions.

Direct searches for new particles at energy-frontier hadron colliders have set increasingly stringent limits on the allowed invariant mass. For example~\cite{Harris:2011bh}, the Super proton-antiproton Synchrotron ($Sp\bar{p}S$) set a lower mass bound on dijet resonances of up to 0.3~TeV depending on the model. This was subsequently extended to 1.3~TeV at the TeVatron.  Since 2010, the Large Hadron Collider (LHC) has probed the highest energy scales ever produced in a laboratory, colliding protons at centre-of-mass energies ($\sqrt{s}$) of 7~TeV and 8~TeV (run 1), 13~TeV (run 2) and currently at a record energy of 13.6~TeV (run 3). This unprecedented energy allows access to new resonances up to ${\cal O}(100)$ times the EW scale, thus exploring a wide range of BSM physics scenarios.

This review summarises LHC searches for high-mass resonances decaying to fermions, focusing on the full run-2 dataset corresponding to an integrated luminosity of $\approx 140$~\ifb. After an overview of the theoretical motivations in section~\ref{sec:models}, the techniques used to search for heavy resonances at ATLAS and CMS are discussed in section~\ref{sec:methods}, highlighting the experimental challenges. Current search results are presented in sections~\ref{sec:leptonic}--\ref{sec:mixed} and the resulting mass reach is summarised in section~\ref{sec:summary_reach}. Other search methodologies and future prospects are surveyed in sections~\ref{sec:other} and \ref{sec:future}, respectively, while section~\ref{sec:conclusion} presents the conclusions.

\section{MOTIVATIONS AND THEORETICAL FRAMEWORK}
\label{sec:models}

This section motivates the necessity for BSM physics, discussing the important issues facing the SM and theoretical models that can address them, with a particular focus on those targeted by the searches considered in the subsequent sections.


\subsection{Additional scalars}
\label{sec:scalar}

As the newest piece of the SM, the Higgs boson is a natural place to search for new physics and many BSM models predict extended scalar sectors.   The most common are two-Higgs doublet models (2HDMs)~\cite{Branco:2011iw}, which postulate an additional Higgs doublet, giving rise to 5 physical states after symmetry breaking: two CP-even Higgs bosons ($h$, usually assumed to be the observed state, and a heavier $H$), a CP-odd Higgs boson ($A$) and a pair of charged Higgs bosons ($H^{\pm}$).  More exotic models, such as Higgs triplets, postulate further bosons.

Several types of 2HDM exist, with the most common being Type I, in which charged fermions only couple to one of the doublets, and Type II, where up-type quarks couple to one doublet while down-type quarks and charged leptons couple to the other.  In each case, the rich phenomenology is described in terms of 6 parameters: 4 physical masses ($m_h, m_H, m_A, m_{H^{\pm}}$), the mixing angle $\alpha$ between the two CP-even bosons and the ratio of the vacuum expectation values $\tan\beta$. In the so-called alignment limit, where $\cos(\beta-\alpha) \simeq 0$, the properties of the lighter CP-even Higgs boson are SM-like.  The best known 2HDM realisation is Supersymmetry (SUSY), which is a Type-II model.  In the Minimal Supersymmetric SM (MSSM)~\cite{Fayet:1976et,Fayet:1977yc}, at tree-level the Higgs-sector properties depend only on two parameters (usually chosen as $m_A$ and $\tan\beta$).  Additional parameters play a role, leading to several benchmark scenarios~\cite{Bagnaschi:2018ofa}, such a $m_h^{125}$ where $m_h \approx 125$~GeV and $m_A \approx m_H$.   An alternative approximate framework is provided by the hMSSM~\cite{Djouadi:2013uqa}, where the SUSY particle spectrum is tuned to provide the radiative corrections necessary to achieve $m_h = 125$~GeV.  


\subsection{New vector bosons}
\label{sec:vector}

Any BSM model that extends the SM gauge group naturally predicts additional vector (spin-1) gauge bosons.  In particular, extensions of the electroweak (EW) sector $SU(2)_L \otimes U(1)_Y$ predict heavier cousins of the SM $W/Z$ bosons, which 
are accessible at the LHC if the breaking of the new symmetry to the SM gauge group occurs at the TeV energy scale. 

The simplest extension consists of an extra singlet $U(1)$ gauge group, $U(1)'$, leading to a heavy neutral boson known as the $\zprime$. Beyond this, there is the possibility of an extra triplet $SU(2)$ group, which gives rise to an additional pair of charged heavy bosons, known as $\wprime^{\pm}$.    An example of a singlet is $E_6$-motivated Grand Unification Theories (GUTs)~\cite{London:1986dk}, which contain two such groups leading to $\zprime_{\chi}$ and $\zprime_{\psi}$ bosons that mix with each other, while triplets include left-right symmetric models (LRSM)~\cite{Mohapatra:1974hk,Senjanovic:1975rk}, which have an $SU(2)_L \otimes SU(2)_R \otimes U(1)_{B-L}$ structure, where $B-L$ is the difference between the
 baryon and lepton number.

A typical, though somewhat artificial, LHC benchmark is the Sequential Standard Model (SSM)~\cite{Altarelli:1989ff}, which predicts new bosons with SM-like couplings to the fermions (including decays to the top-quark ($t$) when kinematically allowed) but no coupling to the SM weak bosons.   The SSM may be extended to include lepton-flavour violating (LFV) couplings~\cite{Murakami:2001cs}.  More generically, the Heavy Vector Triplet (HVT)~\cite{Altarelli:1989ff} is a phenomenological model based on effective field theory (EFT), with a variety of different heavy gauge boson possibilities with parameterised couplings to the SM fermions and bosons.  

Within simplified dark matter models~\cite{Abercrombie:2015wmb}, an additional $U(1)'$ symmetry is augmented by a dark matter particle ($\chi$), in the form of a Dirac fermion that only couples to the new group.   Provided some SM particles are also charged under $U(1)'$, the resultant $\zprime$ can act as a mediator between the SM and DM.  Both vector and axial-vector couplings, with strength $g_{q/\ell}$ to the SM quarks/leptons and $g_{\chi}$ to the DM, are possible. 

\subsection{Extra Dimensions, Gravitons and Quantum Black Holes}
\label{sec:tensor}

The weakness of gravity compared to the other fundamental forces may be explained by extra dimensions (EDs)~\cite{Perez-Lorenzana:2005fzz}, whereby our 3+1 dimensional world is embedded in a higher-dimensional bulk space-time.   Unlike the existing force-carrying bosons, that are constrained to our familiar 4 dimensions, the hypothetical spin-2 particle that mediates gravity, the Graviton ($G$), propagates in the bulk.  The resulting geometrical dilution explains the apparent weakness of gravity compared to the other forces, lowering the true scale of quantum gravity $M_D$ from the Planck scale to the TeV scale.  

Extra-dimensional models can be split into two main types.  The first are models with large extra dimensions (LEDs), such as the Arkani-Hamed-Dimopoulos-Dvali (ADD) model~\cite{Arkani-Hamed:1998jmv}, where the Graviton propagates in $n$ extra dimensions of ${\cal O}(1)$~mm in size.  The other type, typified by the Randall-Sundrum (RS) model~\cite{Randall:1999ee}, contains a single ``warped'' ED, separating our familiar 4D space-time ``brane'' from an additional brane at the Planck scale.  The exponential warping factor allows the required reduction in $M_D$ between the Planck brane and our weak-scale 
brane.  In the original ``RS1'' model only the Graviton can propagate in the 5D bulk, while the ``bulk'' RS extension~\cite{Agashe:2007zd} allows the SM particles to also propagate in the bulk.
In the latter case, the graviton field is required to be localised near the Planck-brane, while the Higgs sector is localised near our weak-scale brane.  This may additionally explain the observed mass hierarchy of the SM particles, since particles that lie closer to the weak brane have a larger Yukawa coupling to the Higgs boson than those further away.   
The presence of standing waves in the ED leads to a tower of Kaluza-Klein excitations for the Graviton ($\gravkk$) in our 4D world that can be searched for at the LHC. 
The coupling is governed by $\kmpl$, where $k$ is the curvature of the warped ED and $\mpl$ is the effective 4D Planck scale.  In addition, the lower $M_D$ leads to the possibility of producing, at the LHC, quantum black holes (QBH)~\cite{Meade:2007sz}. which subsequently decay into two or more particles.

\subsection{Excited fermions}
\label{sec:fstar}

The SM family structure, with three generations of both quarks and leptons, can potentially be explained if the fermions are not actually fundamental but are instead composite~\cite{Harari:1982xy}.  In such models, the fermions are composed of new fundamental constituents, termed preons, bound by a new strong interaction with scale $\Lambda$.   Just like the $\pi$ and the $\rho$ in QCD, a natural consequence of compositeness is the presence of excited quarks ($\exq$) and leptons ($\exl$).  

Excited fermions  ($\exf$) can be produced at the LHC provided they are kinematically accessible~\cite{PhysRevD.42.815}.  In the case of excited quarks, the most obvious mechanism is pair production via $gg$ fusion or $q\bar{q}$ annihilation.  However, the predicted cross section is small and the QCD background overwhelming.  More promising is the production via quark-gluon fusion $qg \to \exq$\footnote{Charge conjugation is implied throughout.}.  In addition, both excited quarks and leptons may be produced via contact interactions (CI): $q\bar{q} \to \exf\bar{q}$ or $q\bar{q} \to \exf\bar{\exf}$. 
The excited fermions subsequently decay either via the CI $\exf \to fq\bar{q}$ or via gauge interactions, such as $\exq \to q g$ or $\exl \to \ell \gamma$. 


\subsection{Vector-like fermions}

A large class of BSM models based on string theory or LEDs predict so-called vector-like fermions~\cite{delAguila:1989rq}.   In contrast to the chiral SM fermions, where only the left-handed (LH) component is charged under the weak-isospin group, vector-like fermions are non-chiral, meaning that left- and right-hand components transform under the same $SU(2)$ representation.  As a consequence, their mass term no longer requires the Brout-Englert-Higgs mechanism to satisfy Gauge invariance, meaning they can obtain a Dirac mass that is not bound to the EW scale and thus evade many of the existing experimental bounds.  


Depending on their colour structure, the new states are either vector-like quarks (VLQs) or vector-like leptons (VLLs).  While many VLQ states are possible~\cite{Aguilar-Saavedra:2013wba}, the simplest extensions give rise to a new up-type $T$ quark and/or down-type $B$ quark, which may be either $SU(2)$ singlets or form a doublet.  Likewise, in the VLL case~\cite{Falkowski:2013jya} one obtains a new charged $\tau'$ lepton, either as a singlet or in a doublet with a neutral $\nu_{\tau}'$.   At the LHC, VLQs can be produced in pairs (via the strong interaction) or singly (via the EW interaction), while VLL are pair-produced.  In both cases, the new heavy fermion states decay to a SM fermion via the exchange of EW gauge or Higgs boson, with a combination of theoretical and experimental arguments motivating preferential coupling to the third-generation fermions.

\subsection{Heavy neutral leptons}
\label{sec:HNL}

The observation of neutrino oscillations implies non-zero neutrino mass, an element not present in the SM, for at least two neutrinos.  Further, if one rejects extremely small Yukawa couplings, the lightness of neutrinos compared to the other fermions suggests a different origin for their mass.  The most common solution is the Seesaw mechanism~\cite{Cai:2017mow}, whereby SM neutrinos are Majorana in nature (i.e.\ their own antiparticles) and acquire a Majorana mass term due to the presence of hypothetical heavy states, known generically as Heavy Neutral Leptons (HNLs), at a scale $\Lambda$. The small SM neutrino mass ($m_{\nu}$) then arises naturally as the result of a suppression via the heavy states: $m_{\nu} \sim y_\nu^2 v^2 / \Lambda$ where $v$ is the vacuum expectation value and $y_{\nu}$ is the neutrino's Yukawa coupling.

Within an EFT context, the new physics can be parameterised at leading order via a lepton-number violating (LNV) dimension-5 operator, known as the ``Weinberg'' operator~\cite{Weinberg:1979sa}, which, after Electroweak Symmetry breaking (EWSB), generates the neutrino's Majorana mass. The LNV nature of the term may drive the baryon asymmetry of the universe via Leptogenisis~\cite{Davidson:2008bu}. 

Restricting ourselves to minimal extensions of the SM fields by a single multiplet, leads to only three possible UV completions of this dimension-5 term at tree level, giving rise to three distinct types of Seesaw mechanism.  The simplest is the Type-I Seesaw, which introduces a heavy gauge-singlet right-handed (RH) neutrino partner for each SM neutrino.   Although ``sterile'', the heavy neutrinos interact with a single generation of SM leptons via a Yukawa coupling, with the RH Majorana mass driving the small SM $m_{\nu}$ after mixing.  Instead of a RH neutrino, the Type-II Seesaw consists of a new Higgs triplet (see section~\ref{sec:scalar}) with the mass of the neutral component providing the new physics scale and leading to a light left-handed (LH) Majorana mass for the SM neutrinos.  Together, Type-I and -II models can be embedded within the LRSM introduced in section~\ref{sec:vector}, giving rise to RH partners of the $W$ and $Z$ bosons in addition to the HNLs.  Finally, the Type-III Seesaw introduces an additional fermionic triplet coupling to the SM gauge bosons.  The resulting SM neutrino mass matrix is similar to Type I, with the mass of the leptons providing the new scale, but features a pair of heavy charged leptons ($L^{\pm}$) in addition to the RH neutrino ($N$).   

\subsection{Leptoquarks}

The symmetry between the quark and lepton sectors of the SM can be naturally explained by a new type of particle that carries both baryon and lepton number, known as a Leptoquarks (LQs), which are predicted by many BSM theories.   LQs can be either scalar (spin-0) or vector (spin-1) particles, each carrying colour and fractional electric charge and thus decaying into a quark and a lepton.  At the LHC, they are produced either in pairs~\cite{Diaz:2017lit} or singly in association with a lepton~\cite{Schmaltz:2018nls}. This results in a wide range of possible final states, which are the subject of an upcoming dedicated review and hence not covered further here. 


\section{ANATOMY OF SEARCHES FOR NEW RESONANCES}
\label{sec:methods}

Searches for heavy BSM resonances at the LHC are predominantly carried out by the general-purpose ATLAS and CMS detectors.   Each detector consists of three main layers: an inner tracking detector, to measure the trajectory of charged particles, surrounded by an electromagnetic and hadronic calorimeter system, to measure the particle's energy, and finally an outer muon detector.  Together, these components provide nearly hermetic coverage of the LHC interaction products.   Both detectors utilise a two-level online trigger system to rapidly identify interesting events to be recorded for offline analysis.  Further details of the detectors can be found in Refs.~\cite{ATLAS:2008xda,CMS:2008xjf}.

While both ATLAS and CMS have performed a plethora of searches for new resonances, probing many possible fermionic final states, the analyses tend to follow variations on a similar strategy, the salient features of which are presented below.  Despite differences in the exact detector configuration and analysis procedure between the two experiments, the resulting BSM resonance sensitivity is broadly similar.

\subsection{Triggering}
\label{sec:trigger}
Before any analysis can begin, the events of interest must be effectively collected amidst the background processes, which are significantly larger in rate before further discrimination, exceeding the available trigger bandwidth and/or offline storage and processing capabilities.  While the exact trigger strategy is optimised for the particular analysis, it can generally be categorised as follows.  Since light charged leptons ($e/\mu$) provide a clean trigger signature, final states containing electrons and/or muons are generally selected using a combination of high transverse momentum ($\pt$) single-lepton triggers and somewhat lower $\pt$ di-lepton triggers.   Final states containing neutrinos, or other undetectable particles, are instead detected by triggers based on large missing transverse energy ($\met$), while hadronically decaying tau leptons tend to be selected either using high-$\pt$ single-tau triggers or $\met$ triggers.   The remaining purely hadronic final states are selected using either a single high-$\pt$ jet trigger or several jets with a large scalar momentum sum.  To maintain an acceptable trigger rate, as both the LHC centre-of-mass energy and the instantaneous luminosity delivered have increased, the thresholds have necessarily increased over time.   The primary run-2 single-light-lepton threshold was maintained around 30~GeV, while the single-tau threshold at the end of the run was in the region of 150~GeV.  The $\met$ trigger threshold was kept just below 100~GeV, while the final single-jet threshold was approximately 500~GeV. 

\subsection{Signal and Background Discrimination}
\label{sec:discrimination}

Once the data are collected, the challenge is to separate potential signal events from the various background processes.  To reduce the background as far as possible while maintaining high signal efficiency, the first step is to  select distinguishing variables upon which selections can be applied. Such variables typically include the multiplicity of final state objects, such as leptons and jets; their kinematics, such as momenta and angular distributions; and global event variables, notably $\met$.  The selections either take the form of a series of cuts on the aforementioned variables or, increasingly, utilise multivariate techniques, primarily Boosted Decision Trees (BDTs) and Neural Networks (NNs), to combine the variables into a single discriminating output variable.  Several signal regions, with different selections, may be defined to provide sensitivity over a broad range of models or parameter space.

The signal and remaining background are compared to the data in order to determine if a signal is present or not.  Rather than simply comparing the predicted signal and background yields to the number of events in the data using a ``cut-and-count'' approach, a final discriminating variable is typically utilised as input to the statistical analysis in order to additionally take into account potential shape differences, thus maximising the sensitivity. 
When searching for a new particle the obvious choice is to perform a ``bump-hunt'',  looking for a peak in the invariant mass of its decay products over a generally smoothly-falling SM background distribution.  If the final state cannot be fully reconstructed due to the presence of undetected particles such as neutrinos, the transverse mass, which neglects the longitudinal momentum, is used as a proxy.   In complex final states, with additional particles or multiple resonances, the discriminating variable is often chosen as a measure of the total energy in the event.  A common example used in several searches is the scalar sum of the $\pt$ of all final state objects ($\HT$), with the signal showing up as a broad excess in the tail of the distribution.   In multivariate analyses (MVAs), the maximum sensitivity is achieved by utilising the MVA output itself as the final discriminant, with any signal giving rise to an excess at high values.  

\subsection{Background Modelling}
\label{sec:bkg}

The backgrounds remaining after selection depend on the specific final state, particularly on the presence or absence of high-$\pt$ charged leptons.  The main backgrounds to final states involving electrons/muons tend to be $V$+jets, $VV$ and $t\bar{t}$ (where $V = W/Z$), with the former dominating in the single- and di-lepton case and the latter two becoming more prevalent in multilepton final states.  In addition, leptonic final states have a smaller background contribution from ``fake'' leptons, i.e.\ non-prompt leptons produced in hadronic decays or jets misidentified as leptons.  This background increases in size as the $\pt$ requirements on those leptons are reduced.  Purely hadronic final states on the other hand are largely dominated by QCD multijet production, particularly at low $\pt$.  The background composition largely dictates the method employed to model the background, both in shape and normalisation.

Processes with leptons are largely modelled via Monte Carlo (MC) simulations based on the highest precision calculations available.  While such simulations generally provide a good description of the inclusive background process, they are often not as reliable in the extreme phase-space tails probed by high-mass resonance searches.   To address this, the MC are usually corrected, at least in normalisation but sometimes also in shape, using data in dedicated control regions (CRs), selected to be kinematically close to the signal region (SR) but depleted in signal and instead enhanced in one or more background process.  The resulting modelling is typically cross-checked in independent validation regions (VRs).  


Unlike the case of leptonic backgrounds, MC simulation is unable to describe the normalisation and shape of background from QCD-initiated mutlijet processes with sufficient accuracy.  Firstly, the precision of the QCD calculations themselves is no match for the statistical power of the large multijet data samples collected at the LHC.   Secondly, there are sizeable uncertainties on both the theoretical (e.g.\ non-perturbative effects and Parton Distribution Functions, PDFs) and experimental side (e.g.\ jet energy scale and resolution).  Since such mis-modellings can potentially mimic a signal, a data-driven background approach is needed instead.  

The most common approach is to parameterise the smoothly-falling background analytically with a functional form that is fit directly to the data spectrum.  This is particularly used in dijet resonance searches~\cite{Harris:2011bh}, which describe the resulting invariant mass spectrum ($\mjj$) by a semi-empirical formula of the general form
\begin{equation}
    f(\mjj) = p_0 (1-x)^{p_1} x^{\sum_{i=2}^{n} p_{i} \ln^{i-2} x}
    \label{eq:dijet}
\end{equation}
where $p_{i}$ are free parameters.  Here, the $(1-x)$ term approximates the PDF behaviour at an average fractional momenta $x = \mjj/\sqrt{s}$ and the $x$ term gives a mass dependence similar to the QCD matrix element.  Different analytic functions, often extensions of equation~\ref{eq:dijet}, are used in various final states.  

The challenge with this method is to include a sufficient number of free parameters to accurately model the background without biasing a potential signal. The optimal number of parameters is often determined by a statistical test, such as a Fischer $F$-test~\cite{lomax2013statistical}, and varied as a systematic uncertainty.  The ability to introduce a spurious signal is tested by a signal+background fit to pseudo-datasets derived from a pure background spectrum, obtained from MC or a data CR, while the potential to absorb a real signal is instead determined by injecting a known signal on top of the background spectrum and performing a background-only fit.  Any spurious signal in the background-only peudo-datasets is included as an additional systematic uncertainty.  

The large uncertainties in the simulation of jet production and hadronisation also make MC unsuitable for modelling fake-lepton backgrounds.  In this case the background is generally determined by measuring a so-called ``fake-factor'' or ``fake-rate'', which represents the probability of a fake lepton to pass the lepton identification requirements of the signal region.   The fake-factor, parameterised as a function of the lepton kinematic quantities, is determined from a CR enriched in fake leptons, after subtracting any residual contributions from real leptons using MC simulation. The fake-lepton background in the SR is then estimated by applying the fake-factor to a distribution of the final discriminating variable obtained from a template region that has the same selection criteria as the SR except with looser lepton identification criteria.  

 
\subsection{Signal Modelling}
\label{sec:signal}

For the analyses presented in this article, the signal is modelled using a dedicated simulation, based on a MC generator that implements the specific BSM process under consideration.  However,  the results of such an approach necessarily have a dependence on the model being studied.  Hence, many analyses aiming at model-independent searches instead model the signal using an analytic parametrisation, generally composed of a Breit-Wigner distribution to model the particle-level mass distribution convoluted with a function such as a Crystal-Ball to account for the detector resolution.  Such an approach allows the intrinsic width of the parametrization to be easily varied in order to present results for different resonance widths.   In doing so, both low-mass off-shell effects and potential interference between the signal and background are usually ignored since they are necessarily model-dependent. 

\subsection{Statistical Analysis}
\label{sec:stats}

The final step in any search for BSM physics is a statistical analysis~\cite{Bohm:2014vmk,Cranmer:2014lly} to assess the compatibility of the selected data with the signal and background models described above.  This is achieved via a maximum-likelihood (ML) fit to the final discriminant using a likelihood composed of Poisson terms describing the total number of events in each bin of the distribution, along with extra ``nuisance parameters'' that characterise systematic uncertainties.

In most of the cases presented in this review a frequentist approach is adopted, whereby a test statistic is constructed using a profile likelihood ratio to discriminate signal- and background-like events.  The nuisance parameters are represented by a set of Gaussian terms centred on the nominal value with the width set to the size of the uncertainty in question, which are usually constrained, or ``profiled``, from the data by including the CRs in the ML fit. A few of the analyses instead use a Bayesian formalism, whereby Bayes's theorem is used to predict the posterior probability density distribution based on one's prior belief.

The first step in the statistical analysis is to ascertain whether the data is compatible with the SM background prediction.  This is done by fitting the data with the background-only model.  The significance of any potential signal is quantified by the local $p$-value, which represents the probability for the background to yield an excess at least as large as the one observed in data.   When scanning for a signal across a range of mass hypotheses there is an increased likelihood of an excess due to a local fluctuation in the data, which is corrected for by a ``trials factor''.  A resulting global $p$-value of $3 \times 10^{-7}$, corresponding to a significance of $5\sigma$, is required to claim an observation or discovery, while $3\sigma$, or a $p$-value of $0.003$, indicates tentative evidence of new physics. 

If no excess is observed, the data are used to set upper limits on the allowed cross section for the signal. This is accomplished by introducing the signal into the likelihood and performing a signal-plus-background fit to the data with a floating signal normalisation to determine the amount of signal that would no longer be compatible with the data.  This is quantified via the confidence level (CL), which is one minus the probability of compatibility, and a 95\% CL is generally required to exclude a signal.  In the so-called frequentist case, the limits are actually evaluated by a modified-frequentist approach known as the ${\text CL}_{\text{s}}$ method~\cite{Read:2002hq}. 

\section{SEARCHES FOR NEW RESONANCES WITH LEPTONIC DECAYS}
\label{sec:leptonic}

This section presents searches for heavy resonances decaying into final states with one, two or more leptons (electrons, muons or taus). The channels under consideration and the analysis references for ATLAS and CMS, as well as summary of lower limits on resonance masses from various BSM models are given in table~\ref{tab:lepton_channels}. The results are detailed in the following sections for the dilepton, single-lepton and multilepton final states.  

\subsection{Dilepton} 

New resonances can decay into a variety of dilepton final states.  First, decays into a same-flavour pair are discussed, with decays into a tau-lepton pair being treated separately to those into a light-lepton pair ($e^+e^-$ or $\mu^+\mu^-$) due to the need for dedicated algorithms to identify hadronic tau decays. These are followed by searches for LFV resonance decays, giving rise to an opposite-flavour lepton pair.

\subsubsection{Dielectron and dimuon channels}
These final states are characterised by a clean and simple experimental signature with excellent detection efficiency and a precise measurement of the dilepton invariant mass. The general analysis strategy follows the one described in section~\ref{sec:methods}, searching for a local excess of signal candidates over a smoothly falling dielectron or dimuon mass spectrum. The search is carried out in a data-driven way, and simulated signal and background processes are only used to determine appropriate functions to fit the data, to study the background compositions and to evaluate the signal efficiency.

Electron and muon candidates are required to pass a set of identification and isolation criteria specifically optimised for high-energy leptons. To remain as model-independent as possible, events with at least two isolated high-$\pt$ electrons or muons are selected, without any veto requirements on additional activity (e.g.\ the presence of jets, additional leptons or $\met$) in the event. 
A key aspect of the analyses is to control the detector response for electrons and muons in the very high invariant mass tail (at TeV scales). The dilepton mass resolutions for events with high-$\pt$ muons and electrons are studied using highly Lorentz-boosted $Z$ boson events.  

\begin{table}[b]
\tabcolsep4.pt
\caption{Summary of the lower mass limits on various BSM models from searches in leptonic final states. Where relevant, the resonance decay under consideration is given in parenthesis.
} 
\label{tab:lepton_channels}
\begin{center}
\resizebox{!}{4.0cm}{
\begin{tabular}{@{}c|c||c|c|c@{}}
 \hline
Final & Reference and  & \multirow{2}{*}{Model} & \multicolumn{2}{c}{Mass Limit (TeV)} \\ \cline{4-5}
State &   Luminosity (fb$^{-1}$) &  &  ATLAS & CMS \\
\hline
 \multirow{3}{*}{ee, $\mu\mu$}  
     & \multirow{3}{*}{\makecell{ATLAS~\cite{ATLAS:2019erb}: 139\\~~~CMS~\cite{CMS:2021ctt}: 140}}   
              & SSM $\zprime$          &  5.1   & 5.2  \\ 
              &   & $\zprime_{\Psi}$   &  4.5   & 4.6  \\ 
              &   & $G$ (\kmpl=0.1)    &    -   & 4.8  \\ 
\hline
\multirow{4}{*}{$\tau\tau$}  
      &  \multirow{2}{*}{\makecell{ATLAS~\cite{ATLAS:2017eiz}: 36.1\\~~~CMS~\cite{CMS:2016xbv}: 2.2}} & \multirow{2}{*}{SSM $\zprime$ }   & \multirow{2}{*}{2.4}  & \multirow{2}{*}{2.1} \\ 
      &   &   &   &  \\ \cline{2-5}
      &  \multirow{2}{*}{\makecell{ATLAS~\cite{ATLAS:2020zms}: 139\\~~~CMS~\cite{CMS:2022goy}: 138}}  &  \multirow{2}{*}{$A,H$}   &   \multirow{2}{*}{1.5 ($\tan\beta$=20) }    &   \multirow{2}{*}{1.2 ($\tan\beta$=20)} \\ 
      &   &   &   &  \\ 
\hline
\multirow{2}{*}{ $e\nu, \mu\nu$  }   
      & \multirow{2}{*}{\makecell{ATLAS~\cite{ATLAS:2019lsy}: 139 \\~~~CMS~\cite{CMS:2022krd}: 138}}   &  \multirow{2}{*}{ SSM $\wprime$ } &   \multirow{2}{*}{6.0} &  \multirow{2}{*}{5.7} \\         
      &   &   &   &  \\ 
\hline
\multirow{2}{*}{$\tau\nu$}        
      & \multirow{2}{*}{\makecell{ATLAS~\cite{ATLAS:2021bjk}: 139\\~~~CMS~\cite{CMS:2022ncp}: 138}}  
      & SSM $\wprime$   & 5.0 & 4.8   \\ 
    & & QBH & - & 6.6 \\ 
\hline
\multirow{4}{*}{ $e\mu, e\tau, \mu\tau$}     
      &  \multirow{4}{*}{\makecell{ATLAS~\cite{ATLAS:2020tre}: 139\\~~~CMS~\cite{CMS:2022fsw}: 138}}  
      & LFV $\zprime (e\mu) $    & 5.0 & 5.0 \\  
  &   & LFV $\zprime (e\tau) $   & 4.0 & 4.3 \\  
  &   & LFV $\zprime (\mu\tau) $ & 3.9 & 4.1 \\  
  &   & QBH  & 5.1-5.9 ($n$=6) & 5.0-5.6 ($n$=4) \\  
\hline
 \multirow{2}{*}{ Multileptons}  &  \multirow{2}{*}{\makecell{ATLAS~\cite{ATLAS:2022pbd}: 139\\~~~CMS~\cite{CMS:2017pet}: 12.9 }} 
        &  LRSM $H^{++}$  &  1.1  & 0.72  \\ 
      & &  LRSM $H^{++} (\tau\tau) $&  -  &  0.54 \\  
 \hline
\end{tabular}
 } 
\end{center}
\end{table}

The shapes of the background contributions from SM processes such as Drell-Yan (DY), pair production of top quarks, as well as single-top-quark and diboson production are estimated from simulation, while backgrounds containing leptons produced inside jets or jets misidentified as leptons, are estimated from control regions in data. 
The MC background template is used to perform fit studies and to decide on a smooth functional form to describe the total background shape. The final functional forms used in ATLAS and CMS are slightly different. 
Concerning the signal, both experiments use a generic line shape described by a non-relativistic Breit–Wigner function convoluted with the detector resolution.

In CMS, the search is performed around the assumed resonance mass in a mass window whose size depends on the assumed intrinsic decay width of the resonance and the mass-dependent detector resolution, allowing the background normalization to be determined from data. 
ATLAS instead uses the invariant-mass sidebands of the expected signal in data to constrain all the fit parameters of the background distribution (the smooth functional form), instead of relying on simulation, leading to a nearly fully data-driven analysis.

No significant data excess above the SM background is observed
(see figure~\ref{fig:dilepton_mass_limit}a) and 95\% CL upper limits are set on the cross section times branching ratio into the corresponding dilepton final state 
(see figure~\ref{fig:dilepton_mass_limit}b).
The limits are interpreted in the context of various models described in section~\ref{sec:models}, and lower mass limits are given in the last two columns of table~\ref{tab:lepton_channels}, being around 5 TeV for superstring-inspired models that predict spin-1 resonances, and 
around 2.5 (5) TeV for spin-2 graviton resonances in the RS model for a coupling parameter $\kmpl = 0.01$ (0.1). 
Lepton flavor universality is tested at the TeV scale by comparing the dimuon and dielectron mass spectra.
The ratio of the dimuon to dielectron differential cross sections as a function of dilepton mass is measured after corrections for detector effects, lepton acceptances, and lepton efficiencies. 
Several uncertainties, the most important one originating from the PDFs, cancel in the flavour ratio. The ratio measurement is observed to be in good agreement with unity (within $\approx 10$\% precision) up $\approx 1.5$~TeV. 


Searches for dilepton resonances are also performed in final states with large $\met$, expected in the case of DM particles produced in association with a new neutral vector boson $Z'$ decaying to same-flavour light leptons. The results are interpreted in the context of several dark-Higgs and light-vector benchmark models~\cite{ATLAS:2023tmv}. 


\begin{figure}[tbh]
    \centering
     \begin{subfigure}[b]{0.49\textwidth}
         \centering
          \includegraphics[height=5.6cm,width=5.6cm]{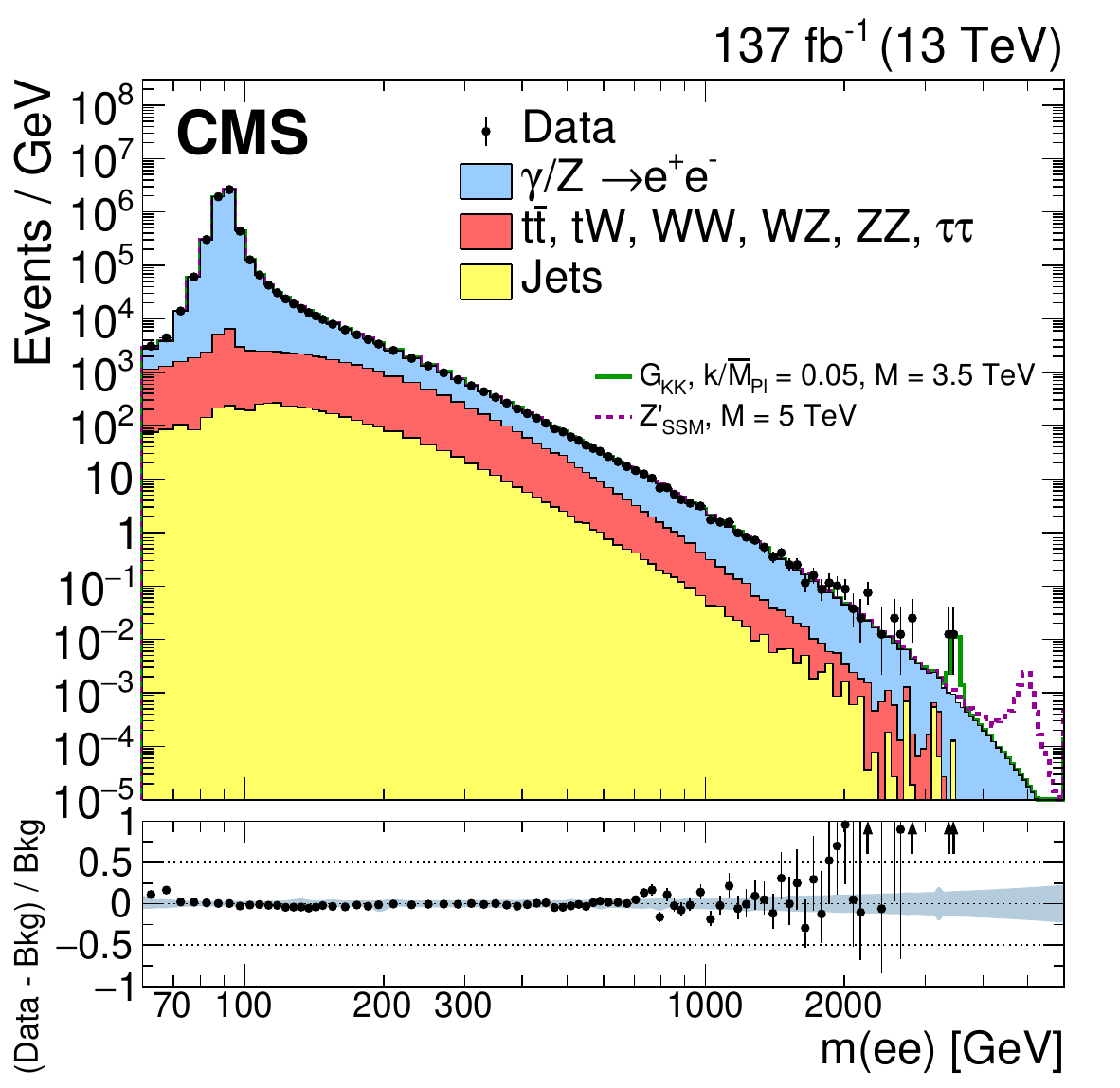}
         \caption{}
         \label{fig:dilepton_Limit}
    \end{subfigure}
    \hfill
     \begin{subfigure}[b]{0.49\textwidth}
         \centering
         \includegraphics[height=5.5cm,width=5.8cm]{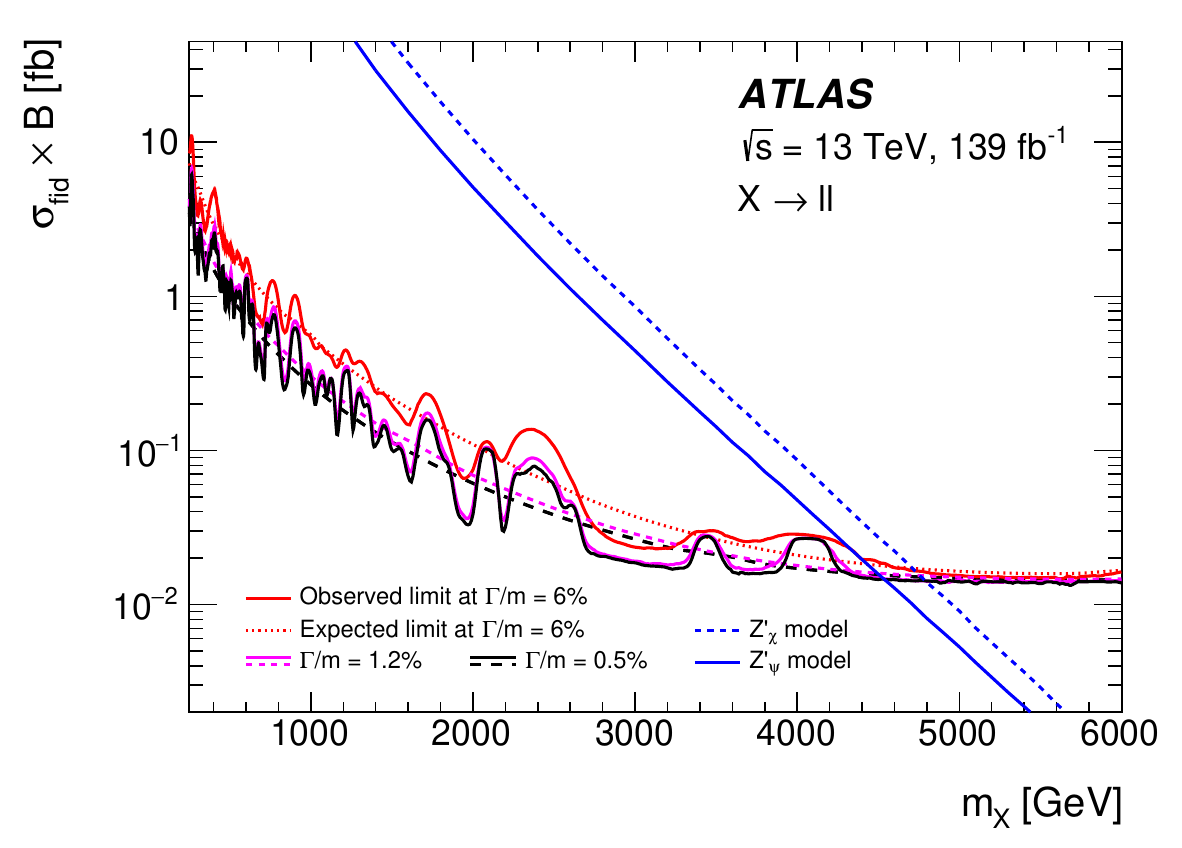}
         \caption{}
         \label{fig:dilepton_Mass}
    \end{subfigure}   
    \caption{Typical mass spectrum and limit plot from dilepton searches: (a) The invariant mass spectrum of the CMS dielectron events (black dots), together with SM expectation (colored histograms) and two signals (a SSM $\zprime$ at 5 TeV and a RS graviton $(\kmpl=0.05)$ at 3.5 TeV)~\cite{CMS:2021ctt}; (b) ATLAS upper limits at 95\% CL on the fiducial cross section times branching ratio as a function of the mass for various signal widths for the combined dilepton channel, together with the theoretical cross sections for Z'$_\chi$ and Z'$_\psi$ in the fiducial region~\cite{ATLAS:2019erb}.}
    \label{fig:dilepton_mass_limit}
\end{figure}

\subsubsection{Ditau channels}
There exist models in which the $\zprime$ boson couples preferentially to third-generation fermions. These are motivated by the high mass of the top quark ~\cite{Lynch:2000md,Malkawi:1996fs} and indications of lepton flavour universality violation in B meson decays~\cite{Faroughy:2016osc}. Using a partial run-2 dataset, CMS and ATLAS have searched for a $\zprime$ bosoon decaying to a tau lepton pair, in $e\mu$, $e\tau_h$, $\mu\tau_h$ or $\tau_h\tau_h$ final states, where $\tau_h$ represents a tau lepton decaying to final states with hadrons and a neutrino. Jets are used as seeds for the $\tau_h$ reconstruction algorithm, selecting those with low track multiplicity (mainly one or three charged pions and possibly up to two neutral pions). The results are interpreted in different models with, for example, the SSM $\zprime$ excluded at 95\% CL for masses below 2.4 TeV. Improved constraints are expected when the entire run-2 dataset is analysed.

The same final state is used to search for neutral (scalar or pseudoscalar) MSSM Higgs bosons in ATLAS and CMS. Both experiments used updated and sophisticated tools to select hadronically-decaying tau leptons.  Indeed, BDT and Deep Neural Net (DNN) algorithms are used by ATLAS and CMS to discriminate $\tau_h$ candidates that originate from genuine tau leptons against $\tau_h$ candidates that originate from quark- or gluon-initiated jets, along with electrons or muons. The DNN inputs include information from reconstructed charged tracks and calorimetric shower shapes in the vicinity of the $\tau_h$ candidate. 
A precise ditau mass reconstruction is important for good separation between signal and background events. However, its reconstruction is challenging due to the presence of neutrinos from the tau-lepton decay.
The discriminating variable used in the analysis is the total transverse mass (using the reconstructed $\pt$ of the two tau-lepton candidates and adding the $\met$ contribution).
Both ATLAS and CMS provide upper limits on the product of the neutral Higgs boson cross section and the branching ratio for the decay into $\tau$ leptons. The results are also interpreted in the context of the m$_h^{125}$ MSSM benchmark scenarios (see section~\ref{sec:scalar}): additional Higgs bosons with masses below 350 GeV are excluded at 95\% CL by CMS, and
values of $\tan\beta > 8$ and $\tan\beta > 21$ are excluded by ATLAS for neutral Higgs boson masses of 1.0 TeV and 1.5 TeV, respectively. \\

\subsubsection{Lepton Flavour Violation channels}

Searches for high-mass resonances undergoing lepton-flavor-violating decays into an electron-muon ($e\mu$) pair, or a electron-tau ($e\tau$) or muon-tau ($\mu\tau$) pair are performed by ATLAS and CMS. The SSM is again used as a benchmark, where the $\zprime$ boson is assumed to have the same quark couplings and chiral structure as the SM $Z$ boson, but allowing for LFV couplings. In final states with $\tau_h$, the ATLAS and CMS analyses are optimised using $\tau_h$ identification based on machine learning techniques as described above.
The SM background in the LFV dilepton search is due to several processes that produce a final state with two different-flavour leptons. For the $e\mu$ channel, the dominant background contributions originate from top-quark pair production and single-top production with the subsequent decay of the $W$ bosons into leptons, along with diboson production. For the  $e\tau$ and $\mu\tau$ channels, multijet and $W$+jets processes are the dominant backgrounds due to the misidentification of jets as leptons. The discriminating variables are the invariant mass distribution for the $e\mu$ channel, and the collinear invariant mass distributions for the $e\tau$ and $\mu\tau$ channels to take into account the presence of the neutrino in the final state. The neutrino four-momentum is reconstructed from the $\met$ and the direction of the visible decay product of the $\tau_h$ candidate.
As no evidence for new physics is observed, lower limits on the mass of a $\zprime$ boson with LFV couplings are set by ATLAS and CMS at values around 5~TeV (4~TeV) for the $e\mu$ ($e\tau$ and $\mu\tau$) final states.  
The analysis results are also interpreted in the context of resonant tau sneutrino production in R-parity violating SUSY models, and in the case of nonresonant quantum black hole (QBH) production in models with $n$ extra spatial dimensions.

\subsection{Single lepton}
As for neutral resonances, searches for heavy charged $\wprime$ resonances have a long history at the LHC in the case of resonances decaying into a charged lepton and a neutrino. As the neutrino is undetected, these events are characterised by a final state with one high $\pt$ charged lepton and large $\met$.
For these searches, the dominant (irreducible) background source originates from the DY production of $W$ bosons. 
Discrimination between signal and background events relies on the transverse mass $m_T$ computed from the charged-lepton $\pt$ and the $\met$ in the event,
$$m_T = \sqrt  { 2 \pt \met ( 1 - \cos{\phi_{\ell \nu}} ) }, $$
where $\phi_{\ell \nu}$ is the angle between the charged lepton and missing transverse momentum directions in the transverse plane.
The analysis is designed to search for the presence of a resonant signal in the high-mass tail of the $m_T$ distribution, where the contributions from background processes are small. 

ATLAS and CMS have analysed the full run-2 dataset, both for light-lepton final states ($e\nu$ and $\mu\nu$) and for the tau final state ($\tau\nu$).
Final results are based on a statistical analysis in which the shape of the signal and both the shape and normalization of the background expectations are derived from MC simulation, except for the background contributions arising from jets misidentified as leptons or from hadron decays.
No significant excess is observed in data above the SM background (see figure~\ref{fig:singlelepton_mass_limit}a), and model-independent upper limits are set on the cross section times branching ratio into the corresponding final state. Within the SSM, the combined results from the electron and muon decay channels exclude a $\wprime$ boson with mass less than about 6 TeV at 95\% CL. 
Limits are derived by CMS on the ratio of $\wprime$ boson coupling strength $g_{W'}$ and the SM weak coupling strength $g_W$. The coupling ratio $g_{W'}$/$g_W$ is unity in the SSM, but various BSM coupling strengths could be possible. Variations in the coupling of the $\wprime$ boson affect its width and consequently its $m_T$ distribution.
ATLAS also provides alternative interpretations in terms of generic resonances with different fixed widths ($\Gamma/m$ between 1\% and 15\%) for possible reinterpretation in the context of other models. 
In the possible case of enhanced coupling to the third generation, it is interesting to search more specifically for $\wprime$ decaying into a tau lepton and a neutrino ($\tau\nu$ final state). In that case, a SSM mass limit up to about 5 TeV in obtained (see figure~\ref{fig:singlelepton_mass_limit}b).

\begin{figure}[t]
    \centering
     \begin{subfigure}[b]{0.49\textwidth}
         \centering
           \includegraphics[height=5.6cm,width=5.6cm]{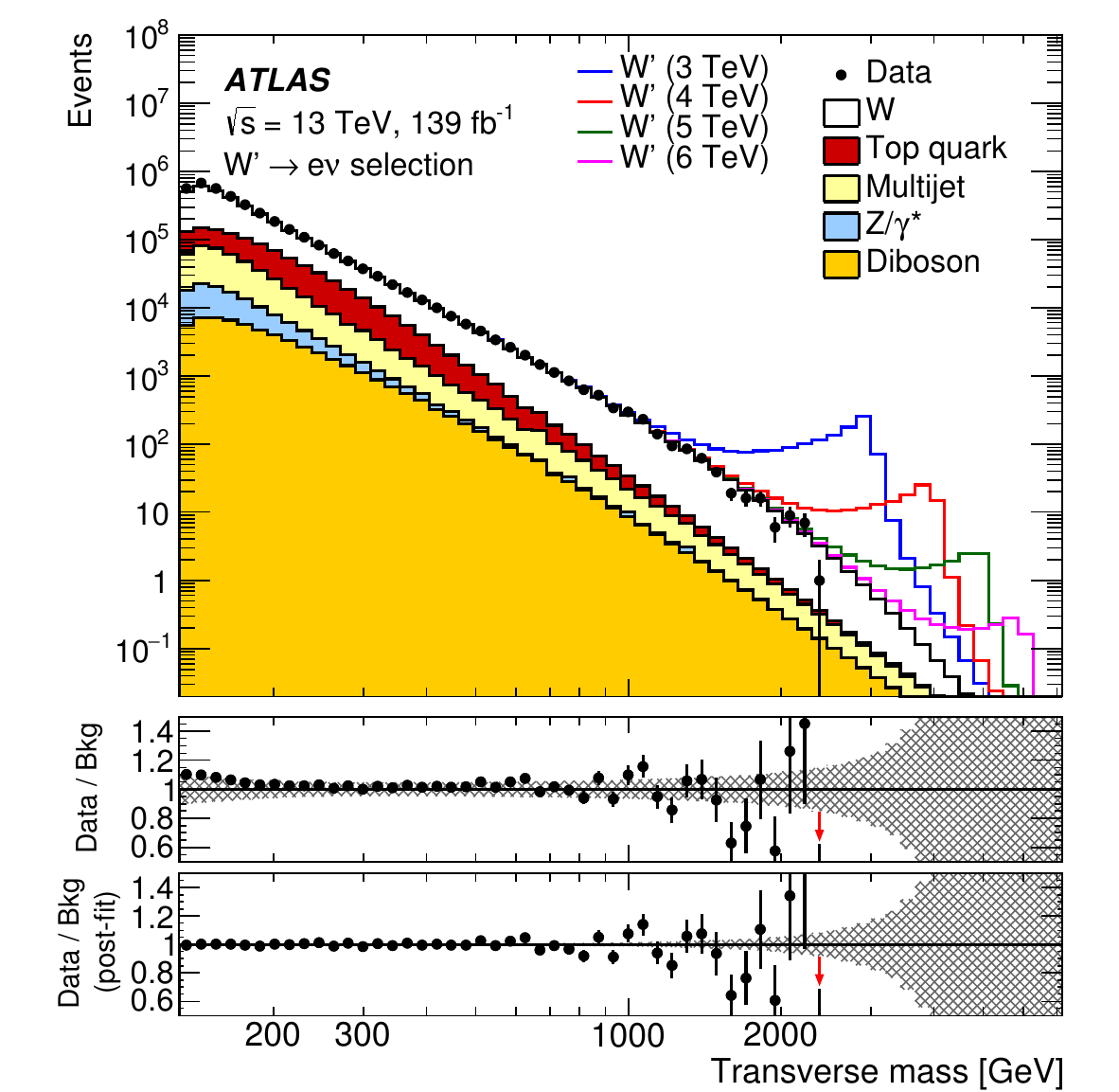}
         \caption{}
         \label{fig:singlelepton_Limit}
    \end{subfigure}
    \hfill
     \begin{subfigure}[b]{0.49\textwidth}
         \centering
         \includegraphics[height=5.7cm,width=5.7cm]{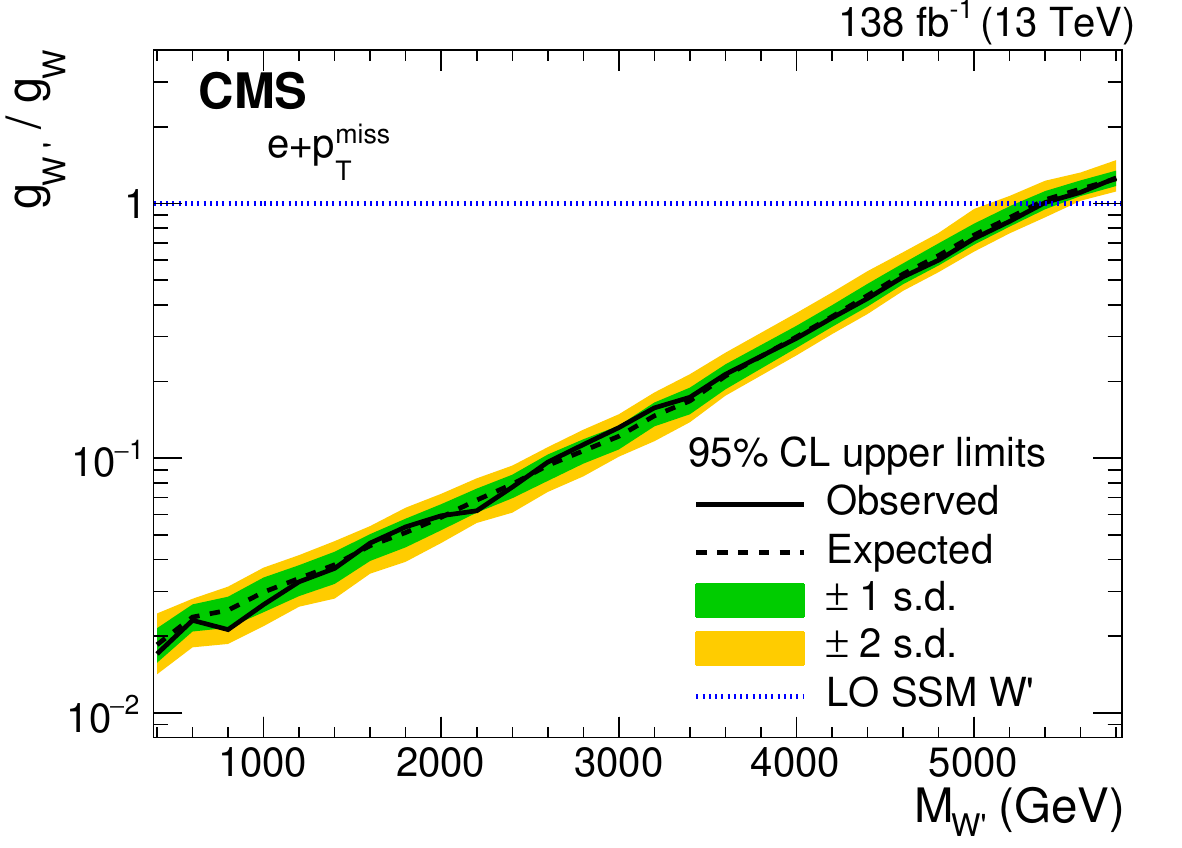}
         \caption{}
         \label{fig:singlelepton_Mass}
    \end{subfigure}   
    \caption{(a) Typical transverse mass spectrum for heavy charged resonance searches, here in the case of ATLAS events for the $e\nu$ decay channel (black dots), together with SM expectation (colored histograms). Expected signal distributions for several SSM $\wprime$ boson masses are shown stacked on top of the total expected background~\cite{ATLAS:2019lsy}; (b) CMS upper exclusion limits on the ratio g$_W'$/g$_W$ for an SSM-like $\wprime$ boson. The unity coupling ratio (blue dotted curve) corresponds to the common SSM benchmark~\cite{CMS:2022krd}.}
    \label{fig:singlelepton_mass_limit}
\end{figure}




\subsection{Multilepton}
Although no doubly-charged bosons are present in the SM, they are predicted in various BSM theories (e.g.\ Type-II seesaw models and the LRSM). Searches were performed for doubly-charged $H$ boson pair production, $H^{++}H^{--}$, or for associated production $H^{\pm \pm}H^{\mp}$, in final states with, respectively, four or three isolated and highly energetic charged leptons. Lower bounds on the $H^{\pm \pm}$ mass have been obtained for a variety of assumptions on its branching ratio to charged lepton pairs.
Considering final states which include electrons or muons, ATLAS obtains a combined lower limit on the mass of the $H^{\pm\pm}$ of 1080 GeV within the LRSM model, assuming equal branching ratio to the 6 charged lepton pair final states. Final states with hadronically decaying tau leptons are also considered by CMS, with a $H^{\pm \pm}$ mass limit of 540 GeV for the case of a 100\% branching ratio to a $\tau\tau$ pair.  


As the exclusive search for new physics did not lead to a discovery, another approach was also followed by ATLAS and CMS in developing generic searches~\cite{ATLAS:2021wob,CMS:2019lwf}. These aim to be sensitive to a wide range of potential new-physics theories simultaneously. Here mutilepton final states (containing at least three charged leptons (electrons or muons)) are considered. The final-state phase space is divided into many event categories according to the lepton multiplicity, $\met$, invariant mass of the leptons, and the presence of leptons originating from a $Z$-boson candidate. In the absence of a detected signal, upper limits are provided in terms of the visible cross sections.
These upper limits can be interpreted in the context of various BSM models as long as the efficiency and acceptance of the respective signals are known. 


\section{SEARCHES FOR NEW RESONANCES WITH HADRONIC DECAYS}
\label{sec:hadronic}

Any BSM state that interacts via the strong force can be produced directly in $pp$ collisions at the LHC and subsequently decay to quarks ($q$) and gluons ($g$). 
Such decays give rise to a generic experimental signature consisting of jets of hadrons, which may optionally be identified as consistent with production from heavy-flavour particles, primarily $b$-hadrons 
or $t$-quarks.

\subsection{Dijet final states}
\label{sec:dijets}

The simplest potential hadronic decay of a new particle consists of the resonant production of a pair of partons.  In the case of a neutral BSM particle this will consist of either a quark -antiquark ($q\Bar{q}$) or two-gluon ($gg$) pair, while in more complex models with fractional charge it may be formed from a quark and a gluon ($qg$).  Regardless of the exact partonic structure, such a decay will give rise to an experimental final state consisting of two energetic jets, whose invariant mass ($\mjj$) forms a resonant peak at the  mass of the new particle.    Consequently, such a final state probes a wide variety of the theoretical models outlined in section~\ref{sec:models}, ranging from new gauge bosons to excited quarks and compositness. 

Since previous hadron colliders have already probed the $\mjj$ region up to around 1.4~TeV~\cite{CDF:2008ieg}, the LHC experiments have primarily focused on searches for resonances with higher masses.   For such searches, the events can be efficiently collected using single-jet triggers without the trigger rate becoming prohibitive.  

In both ATLAS and CMS, the experimental approach to such dijet resonance searches follows the bump-hunting strategy described in section~\ref{sec:discrimination} using the $\mjj$ spectrum as the discriminant.  In CMS, the $\mjj$ resolution is improved by collecting hard-gluon radiation around the leading two jets to form so-called ``wide jets''.   After reducing the QCD background via kinematic cuts on the angular separation between the jets in both $\eta$ and $\phi$, the smoothly falling QCD multijet background is modelled using equation~\ref{eq:dijet}.  

In order to accurately model the $\mjj$ spectrum, CMS has historically split their searches into a high-mass region, using a 3-parameter fit, and a low-mass region, using a more flexible 4-parameter fit.  ATLAS, on the other hand, takes a different ``sliding-window" approach.  Here, rather than attempting to fit the full spectrum at once, the fit is restricted to specific $\mjj$ ranges (the ``windows'') around the signal mass of interest, which are then scanned (or ``slid'') across the spectrum.  An example fit is shown in figure~\ref{fig:dijeta}. This method has the advantage that only 3 parameters are sufficient to fit each window, but requires careful testing to ensure the potential presence of a signal cannot bias the fit.  In their most recent result, CMS replaced the high-mass fit by an alternative data-driven ``ratio method'', whereby the $\mjj$ spectrum is modelled using a template taken from a high $\Delta\eta_{jj}$ CR that is corrected for kinematic difference between the CR and SR using an $\mjj$-dependent transfer factor taken from simulation.  This method provides a more accurate prediction of the background at high mass and is less affected by potential large-width signals.

\begin{figure}[tbh]
    \centering
     \begin{subfigure}[b]{0.49\textwidth}
         \centering
         \includegraphics[height=5.4cm]{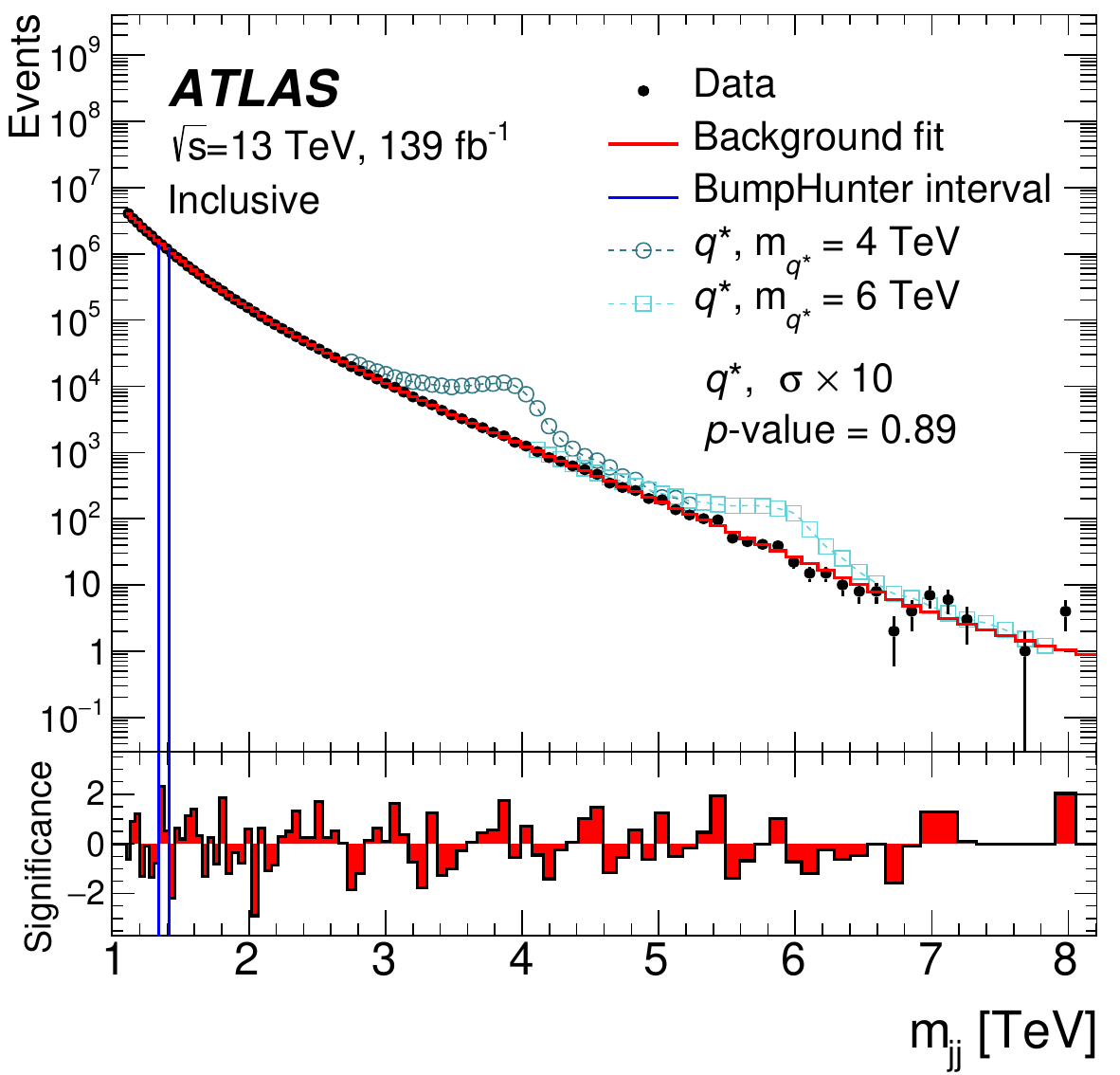}
         \caption{}
         \label{fig:dijeta}
    \end{subfigure}
    \hfill
     \begin{subfigure}[b]{0.49\textwidth}
         \centering
         \includegraphics[height=5.9cm]{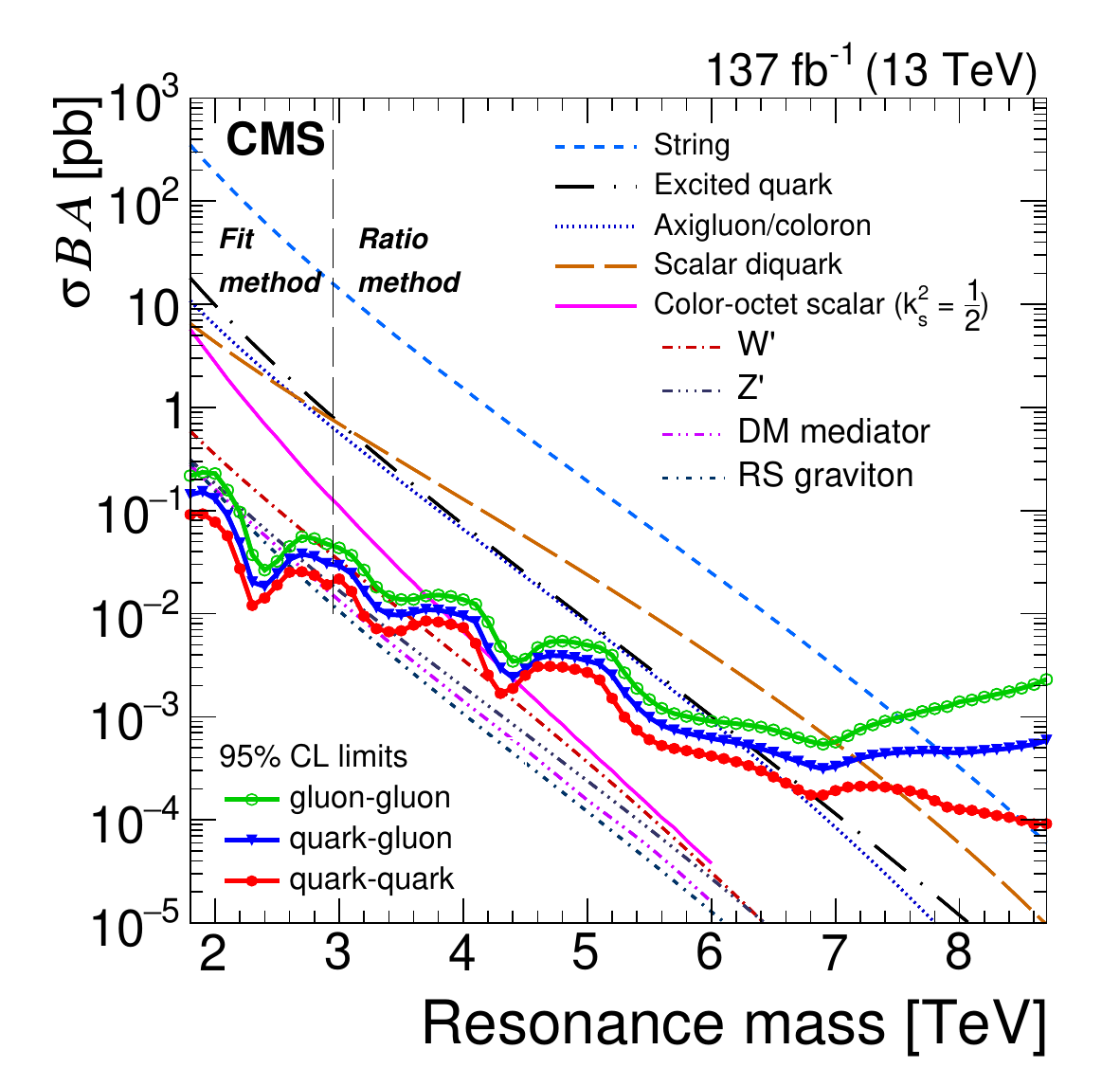}
         \caption{}
         \label{fig:dijetb}
    \end{subfigure}   
    \caption{Example results from dijet searches: (a) ATLAS dijet mass spectrum in data (black points) compared to the background fit (red line) and a hypothetical excited quark signal with a mass of 4 or 6~TeV~\cite{ATLAS:2019fgd}; (b) CMS observed 95\% CL cross section limits for $qq$, $gg$ and $qg$ final states compared to the theoretical predictions for several BSM models~\cite{CMS:2019gwf}.}
    \label{fig:dijet}
\end{figure}

In the absence of a significant excess, limits are set on a variety of BSM models, summarised in table~\ref{tab:hadlimits}.  Beyond the models outlined in section~\ref{sec:models}, CMS sets limits on a range of additional new particles: string excitations of quarks and gluons~\cite{Anchordoqui:2008di,Cullen:2000ef}; scalar diquarks arising from $E_6$ GUTs~\cite{Hewett:1988xc}; axigluons and colourons predicted in extensions of the strong sector~\cite{Frampton:1987dn,Chivukula:2013xla,Simmons:1996fz}; and colour-octet scalars in dynamical EWSB models~\cite{Han:2010rf}.   The resulting lower bounds on the invariant mass range from just under 3~TeV in the case of a SSM $\zprime$ to 7.9~TeV for a string excitation (CMS) or 9.6~TeV for QBHs (ATLAS).  

The results are also used to set model-independent limits on the effective cross section\footnote{cross section times branching ratio times acceptance} for hypothetical resonances with a range of relative widths.  Since the dijet resonance shape depends on the type of the final-state partons, CMS provides separate limits for the $qq$, $gg$ and $qg$ final states using a representative BSM signal simulation in each case, as exemplified in figure~\ref{fig:dijetb}. ATLAS instead assumes a simple Gaussian-shaped signal as a benchmark independent of the parton type.  In both cases, the limit for a narrow resonance ranges from $\mathcal{O}(10^{-1})$~pb at 1--2~TeV to $\mathcal{O}(10^{-4})$~pb at 7--8~TeV.  As the width increases the limits degrade, with ATLAS providing results up to a relative width of 15\% and CMS up to 30\% (55\%) in the case of a spin-2 (spin-1) resonance.

\begin{table}[h]
\caption{Summary of the lower limits on the mass of various BSM resonances from searches in hadronic final states.}
\label{tab:hadlimits}
\begin{center}
\resizebox{!}{4.5cm}{
\begin{tabular}{@{}c|c||c|c|c@{}}
\hline
Final & Reference and & \multirow{2}{*}{Model} & \multicolumn{2}{c}{Mass Limit (TeV)} \\
\cline{4-5}
State & Luminosity (fb$^{-1}$)  & & ATLAS & CMS \\
\hline
\multirow{11}{*}{\makecell{Inclusive\\dijet}} 
& \multirow{11}{*}{\makecell{ATLAS~\cite{ATLAS:2019fgd}: 139\\~~~CMS~\cite{CMS:2019gwf}: 137}}
& SSM $\zprime$ & 2.7 & 2.9 \\
& &  RS $\grav$ & $-$ & 2.6 ($\kmpl = 0.1$) \\
& &  DM $\zprime$ & 3.8 ($g_q = 0.2$) & 2.8 ($g_q = 0.25$)\\
& &  Colour-octet scalar & $-$ & 3.7 \\ 
& &  $\exw$ & 3.9 & $-$ \\
& &  SSM $\wprime$ & 4.0 & 3.6 \\
& &  $\exq$ & 6.7 & 6.3 \\
& &  Axigluon/Colouron & $-$ & 6.6 \\
& &  Scalar $qq$ & $-$ & 7.5 \\
& &  String resonances & $-$ & 7.9 \\
& &  QBH & 9.4 & $-$ \\
\hline
Dijet
& ATLAS~\cite{ATLAS:2019fgd}: 139
& \multirow{2}{*}{$\exb$} & \multirow{2}{*}{3.2} & \multirow{2}{*}{2.5$^{\dagger}$}\\
(1 $b$-tag) & ~~~CMS~\cite{CMS:2022zoc}: 138  & & \\
\hline
\multirow{4}{*}{\makecell{Dijet\\(2 $b$-tag)}} & \multirow{4}{*}{\makecell{ATLAS~\cite{ATLAS:2019fgd}: 139\\~~~CMS~\cite{CMS:2022zoc}: 138}} 
& SSM $\zprime$ & 2.7 & 2.4 \\
& & HVT A & $-$ & 2.4 \\
& & DM $\zprime$ & 2.8 ($g_q = 0.2$) & $-$ \\
& & RS $\grav$  & 2.8 ($\kmpl = 0.2$) & $-$ \\
\hline

4 jet
& \multirow{2}{*}{ATLAS~\cite{ATLAS:2021suo}: 103}
& \multirow{2}{*}{$b$-philic $\zprime$} & \multirow{2}{*}{$-$} & \multirow{2}{*}{1.45}\\
($\ge 3 b$-tag) & & & \\

\hline

$tbtb$ & ATLAS~\cite{ATLAS:2021upq}: 139 
& hMSSM $H^{\pm}$  & $\approx 0.8$ ($\tan\beta=1$) \\

\hline

\multirow{2}{*}{$t\bar{t}t\bar{t}$} 
& \multirow{2}{*}{\makecell{ATLAS~\cite{ATLAS:2022rws}: 139\\~~~CMS~\cite{CMS:2019rvj}: 137}}
& 2HDM $H$  & $\approx 0.49$ ($\tan\beta=1$) & 0.47 ($\tan\beta=1$)\\
& & 2HDM $A$ & $\approx 0.49$ ($\tan\beta=1$) & 0.55 ($\tan\beta=1$)\\

\hline
\end{tabular}
}
\end{center}

\vspace{-0.1cm}

{\footnotesize $^{\dagger}$Including resonant \exb\ production via contact interactions increases this to 4.0~TeV.}

\end{table}

\subsection{Multijet final states}



There have been relatively few general LHC searches for final states with more than 2 jets and this is an area that deserves more attention.  Lately, however, CMS performed the first generic trijet resonance search~\cite{CMS:2023tep} using 138~\ifb.  Data were collected using a combination of high-$\pt$ single-jet triggers and triggers requiring a large $\HT$.  A bump-hunt for a narrow peak in the mass of the three leading wide jets ($\mjjj$), with the background described using a 3-parameter empirical function, was then performed.
The results probe resonance masses from 1.75~TeV to 9~TeV and are interpreted in terms of a $Z_{\text{R}} \to ggg$ resonance.  While the current data are insufficient to constrain this model, limits are set on the effective cross section for narrow three-body resonance decays that range from $\approx 0.1$~fb to $\approx 600$~fb over the mass ranges studied.
The results are also interpreted in terms of cascade decays of a new resonance to $qqq$ or $ggg$.  

Beyond this, both ATLAS~\cite{ATLAS:2023ssk} and CMS~\cite{CMS:2022usq} have searched for a generic resonance $Y$ decaying to two same-mass intermediate states $X$, each of which subsequently decays to a pair of partons, resulting in a 4-jet final state.  Dijet pairs corresponding to the two $X$ states are reconstructed by minimising the $\Delta R$ between the jets and requiring the resultant pair to have a small relative mass asymmetry.  Selected events are categorised into regions where the $X$ decay products have similar Lorentz boost based on $\alpha = \mjjav/\mjjjj$, where $\mjjjj$ is the 4-jet invariant mass and $\mjjav$ is the average dijet invariant mass.  A bump-hunt in then performed on $\mjjjj$ in each category separately, with the background described by a 4-parameter (modified 3-parameter) empirical function for ATLAS (CMS). Both experiments set model-independent limits on the effective cross section for hypothetical narrow $Y$ resonances in various bins of $\alpha$, with ATLAS extending this to relative widths up to 15\%.   CMS sees a modest $3.6\sigma$ local excess for events with $\mjjjj \sim 8$~TeV and $\mjjav \sim 2$~TeV, corresponding to a global significance of $2.5\sigma$, but this is not confirmed by ATLAS.   

\subsection{Final states with $b$-quarks}

Many BSM models predict new particles with large couplings to $b$ quarks, giving rise to potential decays into a $b\bar{b}$ pair or, in the case of excited $b$ quarks, $bg$ pair.  While such resonances again give rise to a dijet final state, the sensitivity can be enhanced by identifying (or ``tagging'') jets containing $b$-flavoured hadrons ($b$ jets).   To do so, both ATLAS and CMS have developed complex multivariate $b$-tagging algorithms that exploit the unique features of $b$ jets, such as tracks that are significantly displaced from the primary interaction vertex and secondary (or even tertiary) vertices, to distinguish them from regular light-flavour jets.
    
The experimental analysis follows the methodology described in section~\ref{sec:dijets}, looking for a bump in the $\mjj$ spectrum, with the only difference being that, depending on the model probed, one or both of the jets must be $b$-tagged.  In all cases, QCD mulijet production remains the dominant background and is parameterised by equation~\ref{eq:dijet} with four parameters for ATLAS and, depending on the data-taking period, three to five parameters for CMS.

ATLAS uses an exclusive 1 $b$-jet SR to search for $\exb$, while CMS uses an inclusive region requiring at least one $b$-tagged jet.  For other models, such as $\zprime$ or $\grav$, ATLAS uses a 2 $b$-jet SR, while CMS uses separate 1 and 2 $b$-jet categories.  An additional category, composed of events where neither jet passes the $b$-tagging requirements but at least one jet contains a muon, is included by CMS to target semi-leptonic $b$-hadron decays and mitigate the loss of $b$-tagging efficiency at high jet $\pt$.  

The limits presented in table~\ref{tab:hadlimits} exclude various BSM resonances with masses up 3.2~TeV depending on the model.  The CMS results are displayed in figure~\ref{fig:dibjet} as an example.  Generic model-independent effective cross section limits on narrow resonances, ranging between $\approx 5$~fb at 1.5~TeV to almost $0.04$~fb at 4.5~TeV in the ATLAS 2 $b$-jet category, are also set.  The muon-in-jet category, which dominates the CMS sensitivity for $\mjj > 5$~TeV, allows them to probe up to 8~TeV, excluding an effective cross section above $\approx 0.2$~fb. 

Due to the vanishing heavy-flavour content of the proton, the above analysis would be insensitive to a $\zprime$ boson that couples exclusively to $b$ quarks.  However, such a resonance may be produced in association with additional $b$ quarks via gluon splitting.   Targeting this, ATLAS performed a bump-hunt in the di-$b$-jet mass spectrum in events with at least four jets, where at least one of the additional jets must be $b$-tagged.  The extra jets enable the use of a trijet trigger to collect data, significantly reducing the jet $\pt$ thresholds and hence probing lower $\mjj$.  However, the asymmetric thresholds of the trijet trigger sculpt the $\mjj$ spectrum preventing the use of the standard empirical formula to model the multijet background.  Instead, a novel functional decomposition~\cite{Edgar:2018irz} method is used to describe the background in terms of a truncated series of orthonormal exponential functions via a process analogous to Fourier analysis.   The result probes $1.3 < \mjj < 3.6$~TeV, excluding a $b$-philic $\zprime$ below 1.45~TeV.


\begin{figure}[tbh]
    \centering
     \begin{subfigure}[b]{0.49\textwidth}
         \centering
         \includegraphics[width=5.8cm]{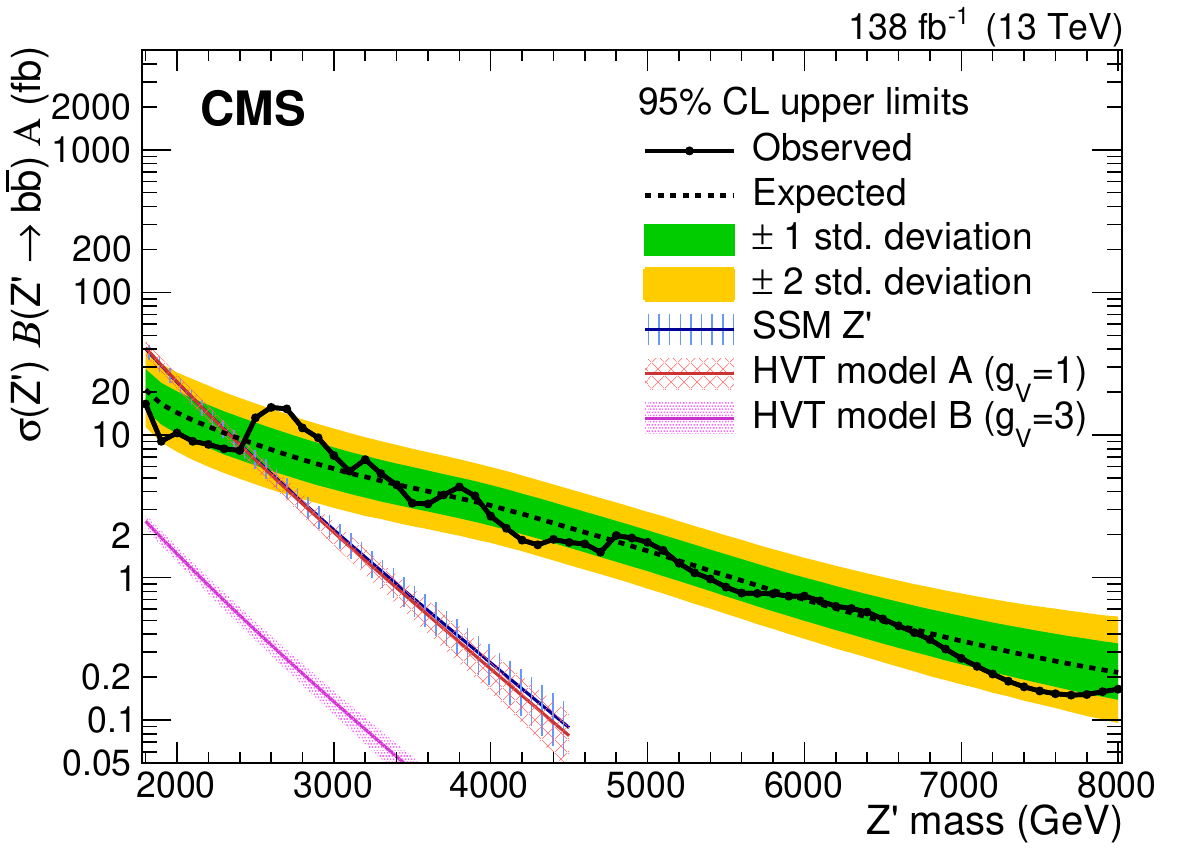}
         \vspace{0.45cm}
         \caption{}
         \label{fig:dibjet}
    \end{subfigure}
    \hfill
     \begin{subfigure}[b]{0.49\textwidth}
         \centering
         \includegraphics[width=5.9cm]{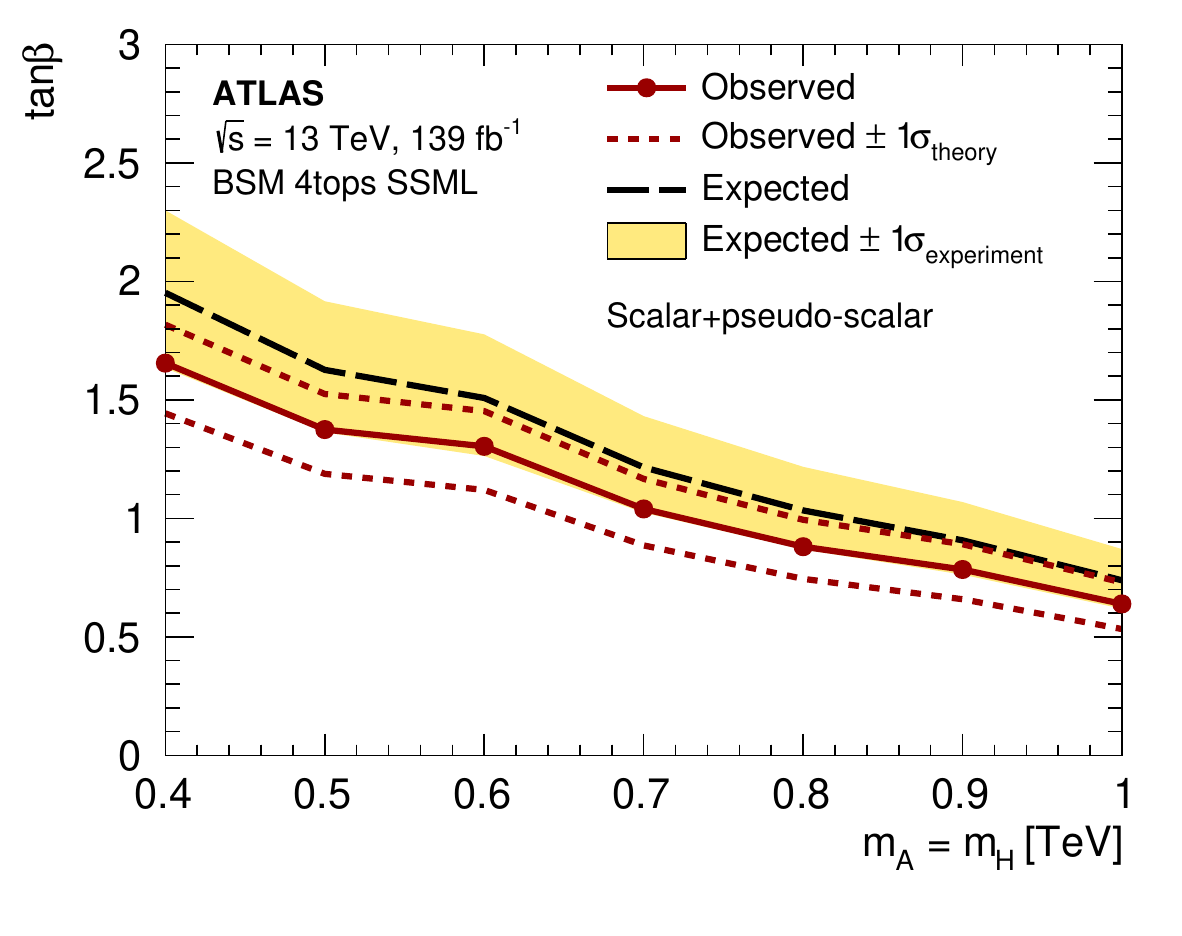}
         \caption{}
         \label{fig:ttbar}
    \end{subfigure}   
    \caption{Examples of searches for resonances decaying to heavy flavours: (a) CMS observed and expected 95\% CL cross section limits for $\zprime \to bb$ compared to the theoretical predictions for two BSM models~\cite{CMS:2022zoc}; (b) ATLAS observed and expected 95\% CL cross section limits for $A/H \to t\bar{t}$ in the Type-II 2HDM~\cite{ATLAS:2022rws}.}
    \label{fig:b_and_top}
\end{figure}

\subsection{Final states with a top quark}

Top-based resonance searches are particularly motivated by the large top Yukawa coupling.  
Hence, the additional Higgs bosons predicted in the extended scalar sectors outlined in section~\ref{sec:scalar} predominantly decay via $A/H \to t\bar{t}$ and $H^{\pm} \to tb$ at high mass.

At the LHC, charged Higgs bosons with $m_{H^{\pm}} > m_t$ are primarily produced in association with a top quark and a bottom quark, before decaying predominantly via $H^{\pm} \to tb$.  ATLAS has performed a full run-2 search for such a signature in the final state with exactly one electron or muon and at least five jets, three or more of which must be $b$-tagged.  The resulting events are categorised into four SRs, based on the number of jets and $b$ jets, and a simulateous fit performed to the output of a NN trained separately in each region.   The dominant $t\bar{t}+\text{jets}$ background is estimated from simulation corrected to data in CRs.   The results set upper limits on the cross section times branching ratio between $3.6$~pb and $\approx 0.02$~pb over the $m_{H^{\pm}}$ range between 200~GeV and 2~TeV.   In the context of the hMSSM and $m_h^{125}$ benchmarks, $\tan\beta$ values in the range 0.5--2.1 are excluded for $m_{H^{\pm}} < 1.2$~TeV.  Mass limits for $\tan\beta = 1$ are presented in table~\ref{tab:hadlimits}.

Inclusive searches for $A/H \to t\bar{t}$ are experimentally difficult due to the large destructive interference with the $gg \to t\bar{t}$ background, which dilutes the resonant peak to a large extent.  Consequently, ATLAS and CMS instead search for $A/H$ produced in association with a pair of top quarks in the $t\bar{t}t\bar{t}$ final state, which is largely free of such interference.  In order to reduce the large potential backgrounds, both experiments target the final state with either two same-sign light leptons (electrons or muons) or at least three leptons with no charge requirement.  The signal-to-background ratio is further enriched by requiring at least two (six) jets in the ATLAS (CMS) case, two or more of which must be $b$-tagged, and large $\HT$.  

The dominant background comes from SM $t\bar{t}t\bar{t}$ production along with $t\bar{t}$ production in association with a vector boson ($t\bar{t}V$) or Higgs boson ($t\bar{t}H$), which is modelled from simulation corrected to data in dedicated CRs.  There is also a significant background from fake leptons and charge mis-identification.   The final discriminant takes the form of a BDT output.  In the CMS case a single BDT is trained to separate both resonant and non-resonant $t\bar{t}t\bar{t}$ events from other backgrounds, while ATLAS uses an additional second BDT, parameterised in terms of the hypothetical resonance mass, to separate the BSM signals.  

Upper limits on the cross section times branching ratio are set separately on the $A$ or the $H$, as well as on their sum under the assumption $m_A = m_H$.  The results are translated into limits on Type-II 2HDMs in the $\tan\beta$ versus $m_{A/H}$ plane, an example of which is shown in figure~\ref{fig:ttbar}, excluding $\tan\beta$ below 1.6 to 0.6 over the mass range from $\approx 2m_t$ to 1~TeV when considering both states.   Mass limits for $\tan\beta = 1$ are again presented in table~\ref{tab:hadlimits}.

Several other BSM theories, ranging from composite Higgs models to bulk RS models, predict high-mass resonances with enhanced couplings to third-generation fermions,  giving rise to further $t\bar{t}$ and $tb$ decays.   Here, the most recent LHC searches have mainly focused on extending the mass reach into previously unexplored territory, leading to top-quarks with a significant Lorentz boost and hence highly-collimated hadronic decay products.   These cannot be resolved using a standard jet-finding algorithm with a typical radius parameter of $R=0.4$ and are instead reconstructed as a single large-$R$ jet, using sophisticated techniques to disentangle the constituent structure.  Such ``boosted'' results are the subject of a dedicated review~\cite{annurev-nucl-102419-055402} and thus not covered here.  Nevertheless, ``resolved'' searches continue to play an important role in probing resonances below around 1.5~TeV, and should be pursued going forward in order to extend the sensitivity to lower cross sections.

\section{SEARCHES FOR NEW RESONANCES WITH MIXED FINAL STATES}
\label{sec:mixed}

Several of the BSM signatures searched for in the previous two sections can also decay into mixed final states, with leptons and jets or fermions and photons, as detailed below. This concerns mainly the search for excited fermions, vector-like fermions, right-handed charged bosons ($\wprimerh$) and QBH.
The final states, the analysis references for ATLAS and CMS, and a summary of lower limits on resonance masses from various BSM models are given in table~\ref{tab:mixed_channels}, and are summarised in the subsections below.

\subsection{Single-lepton $+$ jet(s) final state}
QBHs with masses near $M_D$ can decay into two-particle final states with large branching ratios~\cite{Gingrich:2009hj}. While angular momentum, electric charge, and color are usually assumed to be conserved in strong-gravity interactions, this may not be the case for the baryon or lepton number of the SM. This motivates the ATLAS search for QBHs decaying to one lepton and one quark, looking for an excess of events in the electron+jet and muon+jet invariant mass spectra. Both a model-independent search and a model-dependent one are performed. In the latter, the 5-bin invariant mass distributions of signal and background events in the SRs are fit simultaneously with background events in three control regions (enriched in $W$+jets, $Z$+jets and $t\bar{t}$ events). Upper limits are set on the production cross sections times branching ratios for QBHs decaying into a lepton and a quark, giving a lower mass limit of 9.2 TeV in the ADD model and 6.8 TeV in the RS model.

\subsection{Single-photon $+$ single-jet final state}
Both ATLAS and CMS have searched for BSM signals in events with a photon and a jet.
The main backgrounds to this analysis arise from irreducible SM $\gamma$+jet production (which provides the largest contribution), QCD multijet production, and $W/Z$ + $\gamma$ processes. The analysis strategy consists of searching for a resonance-like excess in the $\gamma$+jet invariant mass  above the expected SM background. A generic search for a Gaussian-shaped signal with various widths (from 2\% to 15 \%) is performed. Two specific models are also considered: excited quarks and QBH (see section~\ref{sec:models}). Excited light-flavor quarks (excited bottom quarks) are excluded up to a mass of 6.0 (3.8) TeV, assuming the compositeness scale $\Lambda$ is equal to the mass of the excited quark. The production of QBHs is excluded for QBH masses up to 7.5 (5.2) TeV in the ADD (RS) model. 

\subsection{Dilepton $+$ photon final state}
Searches for excited leptons decaying into a SM lepton and a photon, $\ell^* \rightarrow \ell \gamma$, provides a clear signature with a high signal selection efficiency. The associated production of an excited lepton, $ \ell \ell^* \rightarrow \ell \ell \gamma$, involves two SM leptons in the final state and therefore there are two possible pairings of a lepton with the photon, corresponding to two invariant masses. A search window is defined in the two-dimensional distribution of these two masses in order to reduce the dominant SM background coming from DY + $\gamma$ events. 
The search performed by CMS is based on a partial run-2 dataset (35.9 fb$^{-1}$). Using a single-bin counting method, excited electrons and muons are excluded for masses below 3.9 and 3.8 TeV, respectively, under the assumption that the excited lepton mass equals the compositeness scale. 

\begin{table}[t]
\caption{Summary of the lower mass limits on various BSM models from searches in mixed final states. Where relevant, the resonance decay under consideration is given in parenthesis.} 
\label{tab:mixed_channels}
\begin{center}
\resizebox{!}{4.0cm}{
\begin{tabular}{@{}c|c||c|c|c@{}}
 \hline
Final & Reference and  & \multirow{2}{*}{Model} & \multicolumn{2}{c}{Mass Limit (TeV)} \\ \cline{4-5}
State & Luminosity (fb$^{-1}$)  & & ATLAS & CMS \\ \hline
$\ell$ + 1 jet   &  ATLAS~\cite{ATLAS:2023vat}: 140 &  ADD(RS) QBH   & 9.2 (6.8)   &   -  \\  \hline
\multirow{3}{*}{$\gamma$ + 1 jet}  
  &  \multirow{3}{*}{\makecell{ATLAS~\cite{ATLAS:2017dpx}: 36.7\\~~~CMS~\cite{CMS:2023twl}: 138}}  
      &  $\exq  $     & 5.3 &  6.0     \\ 
  &   &  $\excitedb$  &  -  &  3.8   \\ 
  &   &  ADD(RS) QBH  &  -  &  7.5 (5.2)   \\ \hline
\multirow{2}{*}{$\ell \ell + \gamma$} &  \multirow{2}{*}{~~~CMS~\cite{CMS:2018sfq}: 35.9 } 
      & $\exel (e \gamma)  $ & - & 3.9 \\ 
  &   & $\exmu (\mu \gamma)$ & - & 3.8  \\  \hline
\multirow{5}{*}{$\ell\ell$ + 2 jets}  
  &  \multirow{2}{*}{\makecell{ATLAS~\cite{ATLAS:2019zfh}: 36.1\\~~~CMS~\cite{CMS:2020cay}: 77.4}}    
      & $\exel (eqq) $     & 4.8 &  5.6 \\ 
  &   & $\exmu (\mu qq)$   & -   &  5.7 \\ \cline{2-5}
  & \multirow{2}{*}{\makecell{ATLAS~\cite{ATLAS:2018dcj}: 36.1\\~~~CMS~\cite{CMS:2021dzb}: 138 }}    
      &  LRSM $\wprimerh (eeqq)$ & 4.7  &  4.7 \\ 
  &   &  LRSM $\wprimerh (\mu\mu qq)$    & 4.7  &  5.0 \\
  &  ~~~CMS~\cite{CMS:2018iye}: 35.9 &  LRSM $\wprimerh (\tau\tau qq)$    & -  &  3.5 \\ \hline
\multirow{2}{*}{$\ell\ell + \ge$ 2 jets}    & \multirow{2}{*}{\makecell{CMS~\cite{CMS:2023ooo}: 138}}
      & LRSM $\zprime (ee qqqq)$ & -  & 3.6 \\ 
  &   & LRSM $\zprime (\mu\mu qqqq)$ & -  & 4.1 \\ \hline
$\tau\tau + \ge$ 2 jets   & ATLAS~\cite{ATLAS:2023kek}: 139  
& $\extau (\tau qq)$   & 4.6 & - \\ \hline
\multirow{4}{*}{Multileptons}  &  \multirow{2}{*}{\makecell{ATLAS~\cite{ATLAS:2020wop,ATLAS:2022yhd}: 139\\~~~CMS~\cite{CMS:2022nty}: 138}}   
    &  \multirow{2}{*}{Type-III seesaw $L^{\pm}$} &  \multirow{2}{*}{0.91} & \multirow{2}{*}{0.98} \\ 
    &  &   &  &  \\ \cline{2-5}
    &  \multirow{2}{*}{\makecell{ATLAS~\cite{ATLAS:2023sbu}: 139\\~~~CMS~\cite{CMS:2022nty}: 138}}   
    &  \multirow{2}{*}{VLL $\tau'$} &  \multirow{2}{*}{0.90} &   \multirow{2}{*}{1.0}\\ 
    &  &   &  &  \\ 
\hline
\end{tabular}
} 
\end{center}
\end{table}

\subsection{Dilepton $+$ jets final state}
Another way to search for excited leptons is via their decay into a lepton and a pair of quarks, $\ell^* \rightarrow  \ell q \Bar{q}$, where both the production and the decay of the excited leptons occur via a contact interaction with a characteristic energy scale $\Lambda$. The branching ratio of this decay mode increases with the mass of the $\exl$, providing the most sensitive channel for very heavy excited leptons. 
CMS has searched for excited electrons and muons in the $e(ejj)$ and $ \mu(\mu jj)$ final states using a partial run-2 dataset (77 fb$^{-1}$), looking for an excess of events in the invariant mass distribution of the two leptons and the two highest-$\pt$ jets. The main background arises mainly from DY and $t\bar{t}$ events, with a smaller contribution from single-top, multiboson and $W$ + jets events.
ATLAS has concentrated on the electron channel, considering also the $\exel$ decay via a gauge interaction into a neutrino and a $W$ boson, using 36.1 fb$^{-1}$ of data. Upper limits are calculated on the production cross sections as a function of the excited lepton mass, excluding excited electrons and muons with masses below 5.6 and 5.7 TeV, respectively, at 95\% CL.
ATLAS also performed a search for excited tau leptons in events with two hadronically-decaying tau leptons and two or more jets, using the full run-2 dataset. 
The discriminating variable used in the analysis is the scalar sum of transverse momenta of the two jets and the two tau leptons, referred to as $S_T$. The signal would appear as an excess of events in the tail of the $S_T$ distribution, as shown in figure~\ref{fig:mixed}a. No signal is observed and, in the case where $\Lambda$ is equal to the excited tau-lepton mass, excited tau-leptons with masses below 4.6 TeV are excluded. The results are also interpreted in the context of a search for LQs.

The final state consisting of two same-flavor leptons and two jets is also considered to search for a heavy $\wprimerh$ and a heavy right-handed neutrino ($N_R$) predicted by the LRSM (see section~\ref{sec:HNL}).
The dominant production process for the $\wprimerh$ boson at the LHC is via the DY mechanism, $q\Bar{q'} \rightarrow \wprimerh \rightarrow \ell N_R$, the $N_R$ subsequently decaying into a lepton and a pair of quarks. Hence, the expected signal is characterised by an excess of events in the invariant mass distribution of the four final-state objects. The potential Majorana nature of right-handed neutrinos implies that the final-state charged leptons can have the same sign. 
ATLAS and CMS have performed a search in the case of light leptons (electrons and muons). The CMS search covers two regions of phase space, one where the decay products of the heavy neutrino are merged into a single large-area jet (boosted topology), and one where the decay products are well separated (resolved topologoy). The addition of the boosted topology improves the search sensitivity for low $N_R$ masses (below 0.5 TeV). 
No signal is observed and lower limits are set on the masses in the right-handed boson and neutrino mass plane. For $N_R$ masses equal to half the $\wprimerh$ mass, $\wprimerh$ masses are excluded up to 4.7 and 5.0 TeV for the electron and muon channels, respectively. The CMS results for the muon channel are presented in figure~\ref{fig:mixed}b.

CMS has extended the search to the case of final states with two tau leptons that decay hadronically and at least two energetic jets, using the 2016 dataset (35.9~\ifb). Assuming that the $N_R(\tau)$ mass is half of the $\wprimerh$ mass, masses of the $\wprimerh$ boson below 3.5 TeV are excluded at 95\% CL. 

In the LRSM context, a search for pair production of $N_R$ through an extra neutral gauge boson $\zprime$ is also performed by CMS in the case where each $N_R$ decays to a lepton and two quarks, leading to a final state with two leptons and four quarks. The kinematic distributions of the final state are strongly dependent on the ratio of the $N_R$ mass and the $\zprime$ mass. Three signal regions are defined, depending on the number of small- and large-radius jets, arising from resolved and boosted topologies respectively. The signature of a signal is an excess of events in the invariant mass distribution of the final-state objects, two same-flavor leptons (e or $\mu$) and at least two jets. 
For a $N_R$ mass equal to one quarter of the $\zprime$ mass, the observed 95\% CL lower limit on the mass of the $\zprime$ boson is 3.6 (4.1) TeV in the dielectron (dimuon) channel. Dedicated algorithms are optimised for boosted signals providing a significant improvement in sensitivity in the case of low $N_R$ mass. For an $N_R$ mass of 100 GeV, the $\zprime$ mass limits are 2.8 and 4.4~TeV for the two channels, respectively.

\begin{figure}[tbh]
    \centering
     \begin{subfigure}[b]{0.45\textwidth}
         \centering
          \includegraphics[height=5.0cm,width=5.5cm]{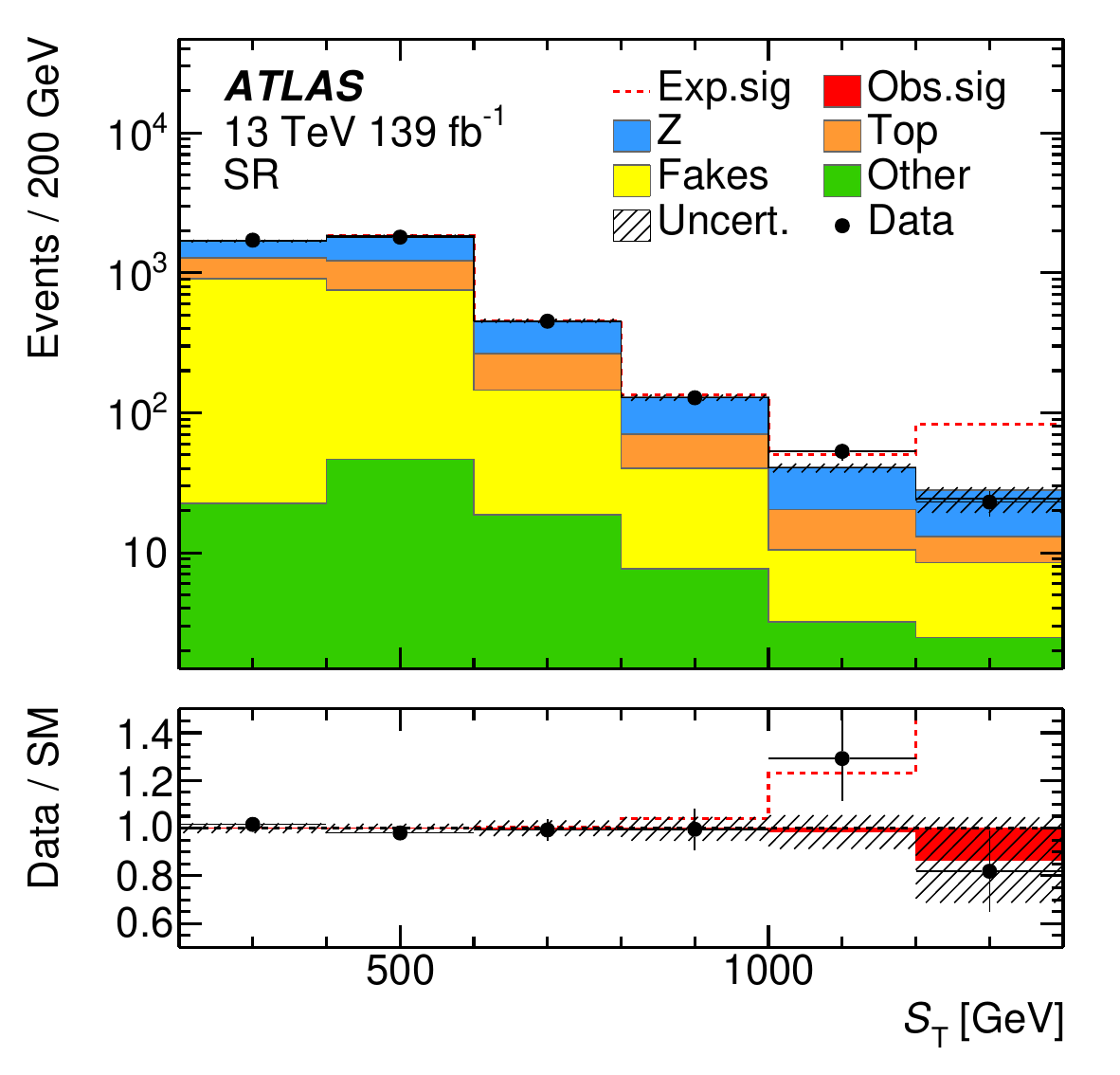}

         \caption{}
         \label{fig:mixed_ATLAS}
    \end{subfigure}
    \hfill
     \begin{subfigure}[b]{0.50\textwidth}
         \centering
         \includegraphics[height=5.0cm,width=6.0cm]{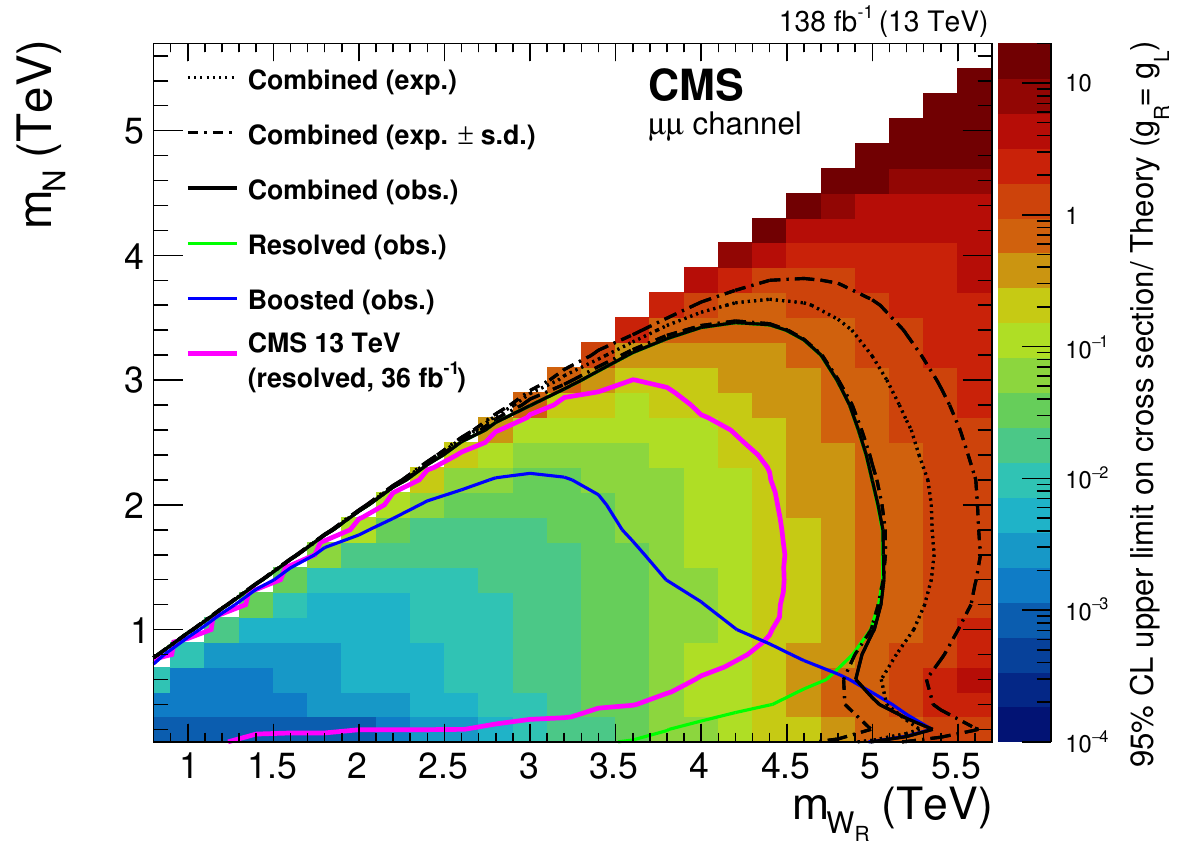}
         \caption{}
         \label{fig:mixed_CMS}
    \end{subfigure}   
    \caption{Examples of searches with mixed (leptons and jets) final states: 
     (a) Distribution of the $S_T$ variable for the ATLAS data and the SM backgrounds, an excited tau signal (in red) at a mass of 1500 GeV is also shown~\cite{ATLAS:2023kek};    
    (b) CMS upper limits on the product of the production cross sections and the $\wprimerh$ branching ratio divided by the theory expectation 
    for the muon channel, in the $\wprimerh$ and $N_R$ mass plane~\cite{CMS:2021dzb}.
    }
    \label{fig:mixed}
\end{figure}

Should a new $\zprime$ boson couple preferentially to second- or third-generation quarks, it is interesting to search for high-mass dimuon resonance production with an explicit requirement on the presence of $b$ jets. Such an analysis has been performed by CMS~\cite{CMS:2023byi}, in which the events are categorised according to the $b$-jet multiplicity. The dominant backgrounds arise from DY processes and $t\bar{t}$ production. The DY background is significantly reduced by requiring the detection of at least one $b$ jet. To suppress the $t\bar{t}$ background, the minimum invariant mass of any muon plus $b$ jet pair is required to be larger than the top-quark mass. The background is estimated directly from the data across the full dimuon mass range. Model-independent limits are derived on the number of signal events with exactly one or more than one $b$ jet. 
Results are also interpreted in a lepton-flavor-universal model with $\zprime$ coupling to $b\Bar{b}$ or $b\Bar{s}$, and to leptons, where the $\zprime$ couplings to all neutrinos and to all charged leptons are assumed to be equal.

\subsection{Multilepton $+$ jets final state}

Inclusive nonresonant multilepton final states are also used to probe potential new physics. In this context, ATLAS and CMS analysed final states with two or more leptons of different flavour and charge combinations, with potential additional jets and $\met$.

A search for the pair production of heavy leptons predicted by the Type-III seesaw mechanism (see section~\ref{sec:HNL}) is performed : 
the production of a neutral Majorana and a charged heavy lepton ($N L^{\pm}$) or a pair of charged heavy leptons ($L^{\pm} L^{\mp}$), both coming from the $s$-channel exchange of virtual EW gauge bosons. The heavy leptons subsequently decay to a SM lepton and a EW gauge boson or a Higgs boson.
ATLAS presented two analyses, one focused on the dilepton final state and the other on multilepton (greater than two) final states, considering electrons and muons (including those from leptonic tau decays). CMS considered final states with three or more charged leptons, including hadronically-decaying tau leptons. 
Lower limits are set on the mass of heavy fermions in the range 845–1065 GeV for various decay branching ratio combinations to SM leptons. In the case of flavour-democratic mixings with SM leptons, the ATLAS and CMS limits are 0.91 and 0.98 TeV, respectively.

In the CMS analysis, other BSM scenarios such as vector-like tau lepton and scalar LQs are also probed. Here, the events are categorised based on the lepton and $b$-jet multiplicities and various kinematic variables, before a signal-specific BDT is used to enhance the sensitivity.  
Doublet and singlet vector-like tau lepton extensions of the SM are excluded for masses below 1045 GeV and in the mass range 125–150 GeV, respectively. 

ATLAS has performed a dedicated search for vector-like leptons coupling to third-generation SM leptons in the multilepton final state with zero or more hadronic tau lepton decays. To maximize the separation of signal and background events, a MVA discriminant was used. In the context of a doublet model, vector-like leptons coupling to third-generation SM leptons are excluded for masses below 900 GeV at the 95\% CL.

\section{SUMMARY OF PHYSICS REACH}
\label{sec:summary_reach}


The previous sections present the various LHC searches for heavy resonances in terms of  their signature,  grouped into leptonic final states, hadronic final states and mixtures of both. The leptonic and/or hadronic content of the final states defines the experimental approach, including the  selection strategy (in particular the triggers chosen, the object reconstruction and the selection optimisation), the background composition and the estimation of the systematic uncertainties. This section brings the results in different final states together in terms of the BSM models described in Section~\ref{sec:models}, summarising the mass reach for a selection of channels in figure~\ref{fig:summaryPlot}.  In many categories of models, specifically additional gauge bosons, excited fermions and extra dimensions, the LHC is already probing the multi-TeV mass range. In the SSM benchmark,  $\zprime$ and $\wprime$ boson masses are excluded up to about 6 TeV, the leptonic decay channels providing the strongest limits, while the mass reach for excited leptons is also approaching the 6-TeV range. The most stringent exclusion is for QBHs, reaching almost 10 TeV in the dijet channel. Concerning instead additional scalars, heavy neutral leptons and vector-like leptons, the reach is considerably lower, probing masses in the 1-TeV range. For example, additional neutral scalars (vector-like leptons) are excluded up to 1.5~TeV (1.0~TeV) in the tau channel.

\begin{figure}[t]
    \centering
    \vspace{5.5cm}
    \includegraphics[angle=90,width=12.8cm]{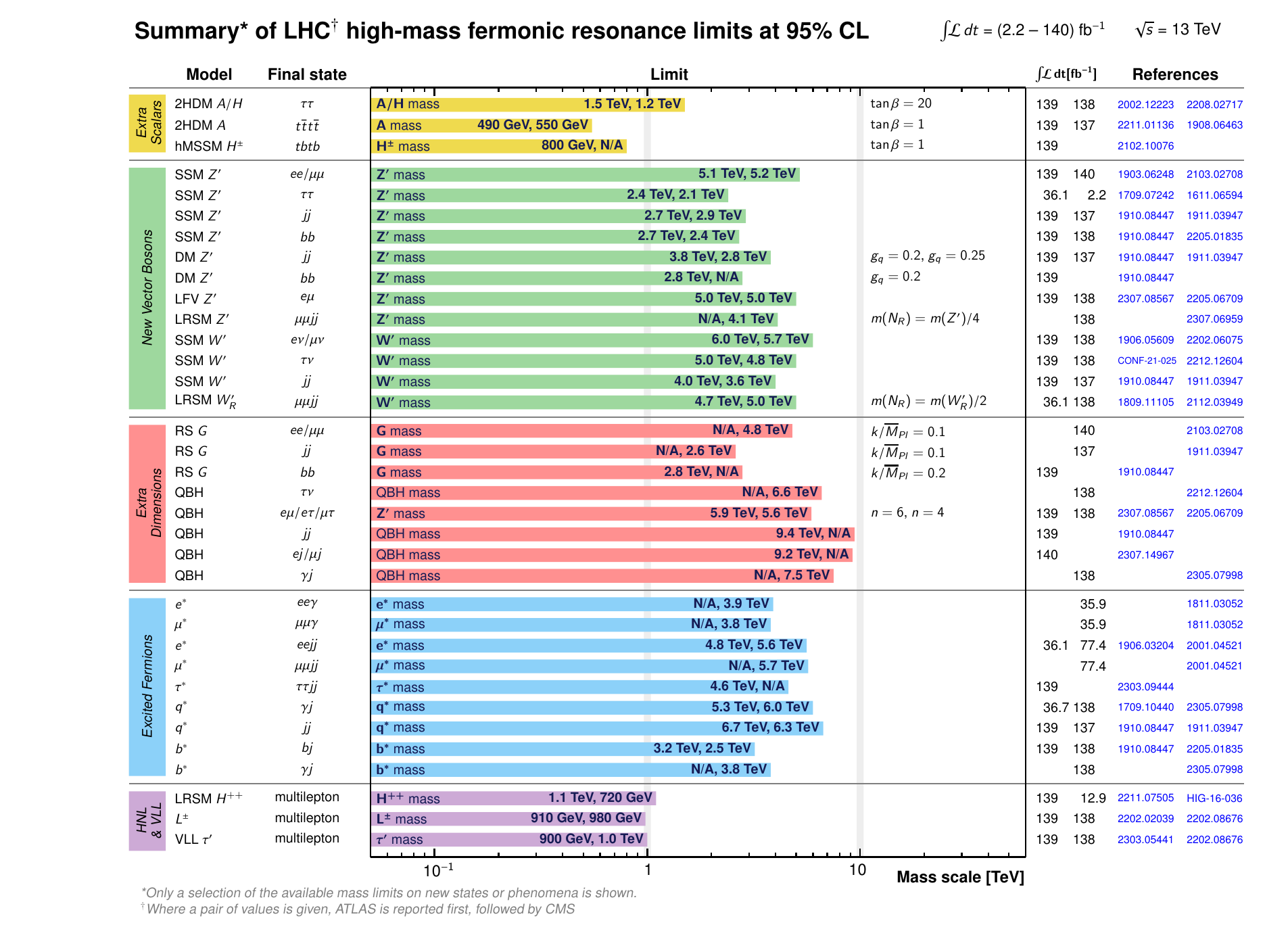}
    \vspace{4.5cm}
    \caption{A summary of the LHC mass reach for high-mass fermionic resonance searches in selected channels. Where a pair of values is given ATLAS is reported first, followed by CMS.  Each bar displays the higher of the two limits. The figure was produced using a modification on the ATLAS summary plot code.}
        \label{fig:summaryPlot}
\end{figure}         

\section{OTHER METHODOLOGIES FOR HIGH-MASS RESONANCE SEARCHES }
\label{sec:other}

While the LHC experiments have, understandably, focused primarily on pushing searches for fermionic resonances to ever higher masses, probing previously inaccessible phase space, it is also interesting to search for lower-mass resonances with smaller cross sections that would have evaded previous collider searches.   The key challenge in accessing the low-mass region at the LHC is triggering effectively while keeping the rate at a manageable level.  There are two main strategies to achieve this.   The first is to perform a so-called ``trigger-level'' analysis~\cite{ATL-DAQ-PUB-2017-003, Mukherjee:2766071}, whereby a lower trigger threshold is accommodated without exceeding the rate limitation by storing only a subset of the full event information and/or performing the analysis directly on trigger objects, avoiding the need to process the full data through the offline reconstruction.   This has allowed both ATLAS and CMS to record jet data with a significantly reduced $\pt$ threshold, below 100~GeV, and hence to probe dijet resonances down to $\mjj \sim 0.5$~TeV~\cite{ATLAS:2018qto,CMS:2016gsl}.  CMS has also used this technique to search for dimuon resonances with masses bewteen 10 and 200 GeV~\cite{CMS:2019buh}, or even lower~\cite{CMS:2023hwl}. The second is to trigger on an additional particle, such as a photon from initial-state radiation (ISR) ~\cite{ATLAS:2019itm,CMS:2019xai} or, in the case of hadronic final states, an associated lepton~\cite{ATLAS:2020zzb}.   

At the other end of the mass scale, direct resonance searches are constrained to below the LHC centre-of-mass energy.  However, resonances with mass above this can still be searched for indirectly via off-shell effects, affecting the kinematics of the decay products.   For example, ATLAS and CMS have performed searches for non-resonant effects in the angular spectrum of both dijet~\cite{ATLAS:2017eqx,CMS:2018ucw} and dilepton ($ee$ or $\mu\mu$)~\cite{ATLAS:2020yat,CMS:2021ctt} pairs, setting lower limits on contact interactions up to a scale of almost 30~TeV and 40~TeV, respectively.

All the searches reviewed above assumed that the new particle has a negligible lifetime, such that the decay products are produced promptly in the detector.   However, particles with longer lifetimes occur naturally when the matrix element of the underlying process is small, for example due to an approximate symmetry or highly virtual intermediate state, and/or there is limited phase space for the decay, perhaps due to a mass degeneracy~\cite{Knapen:2022afb}.  Indeed, the SM provides many historical precedents for particles with macroscopic lifetimes and many BSM theories predict new particles with a range of lifetimes.   Depending on the lifetime, such long-lived particles (LLPs) lead to a wide variety of experimental signatures, including displaced vertices and trackless jets amongst others.  The lack of BSM physics observations have only increased the interest in LLPs in recent years and spurred the development of novel techniques to exploit ATLAS and CMS to search for them.  Such analyses are beyond our scope and are the subject of several dedicated reviews~\cite{Knapen:2022afb, Alimena:2019zri}.  Nevertheless, it is important to mention that closing the gap between promptly decaying and long-lived particles is an import challenge for future searches.

Finally, the fermionic resonance searches reviewed here are naturally complemented by searches for resonances decaying into various bosonic final states.  A summary of diboson searches, for example, can be found in Ref.~\cite{Dorigo:2018cbl}.


\section{PROSPECTS AT THE HL-LHC AND FUTURE COLLIDERS}
\label{sec:future}

As of the writing of the current review, LHC run-3 is ongoing and will continue operation until the end of 2025, more than doubling the run-2 integrated luminosity.  With 300~\ifb\ the LHC dijet mass reach will extend to $\approx 10.5$~TeV depending on the model~\cite{Chekanov:2017pnx}, while ATLAS  is predicted to exclude a SSM $\zprime \to \ell\ell$ ($\wprime \to \ell\nu$) with mass up to around 5.4~TeV (6.7~TeV)~\cite{ATL-PHYS-PUB-2018-044}.

Following a long shutdown to upgrade the accelerator complex and experimental detectors, the high-luminosity phase of the LHC (HL-LHC) is scheduled to start in 2029.  Over a decade or more it will collect 3 \iab, 10 times the phase-1 dataset, significantly extending the reach for new resonances~\cite{CidVidal:2018eel}.  The projected $5\sigma$ discovery reach for an SSM $\zprime$ ($\wprime$) will reach masses up to 6.4~TeV (7.7~TeV) in the $\ell\ell$ ($\ell\nu$) final state.  In final states with $\tau$ leptons, the equivalent $\wprime$ discovery prediction is 7.0~TeV, while for the $H/A \to \tau\tau$ it reaches 2.5~TeV at $\tan\beta=50$.  If new resonances are not discovered, the corresponding mass exclusion limits will be increased by a factor of 1.3-1.5.  In the case of excited electrons or muons, the forecast discovery (exclusion) reach is 5.1~TeV (5.8~TeV) in the $\ell\gamma$ decay mode.  For hadronic resonances, the estimated dijet mass reach is 11.2~TeV in the inclusive channel and around 9.5~TeV in the di-$b$-jet channel.  In final states with top-quarks, the discovery reach for a $\wprimerh \to tb$ in the semi-leptonic decay mode is 4.9~TeV, while for a RS $g_{kk} \to t\bar{t}$ it is 6.6~TeV when combining the semi-leptonic and all-hadronic channels and utilising substructure techniques.  In addition, the increased luminosity will enable a significant reduction in the effective cross section upper limits over the entire mass range.

In the longer term, future colliders open the door to probe a significantly wider parameter space for BSM resonances.   New lepton colliders, circular (FCC-ee) or linear (ILC/CLIC), naturally have a much lower background rate, allowing to probe notably smaller BSM cross sections within the existing mass reach (e.g.\ FCC-ee provides the best potential sensitivity for GeV-scale HNLs~\cite{Bose:2022obr}).   At the higher proposed centre-of-mass energies, linear colliders can even increase the mass reach with, for example, a 1~TeV ILC providing SSM $\zprime$ discovery reach up to 14~TeV~\cite{Bose:2022obr}. Beyond this, the proposed future circular hadron collider (FCC-hh) at a 100~TeV centre-of-mass energy will markedly increase the mass reach~\cite{Helsens:2019bfw}.  For a SSM $\zprime$, the projected FCC-hh discovery reach is above 40~TeV in the $\ell\ell$ channel, extending the HL-LHC reach by more than a factor of 6, and approaching 20~TeV in both the $\tau\tau$ and $t\bar{t}$ channels.  The corresponding dijet mass reach is roughly 4 times larger than the HL-LHC, reaching 40~TeV in the excited quark case.




\section{CONCLUSIONS}
\label{sec:conclusion}

The search for new massive resonances decaying to fermions is a mainstay of the BSM physics program of the ATLAS and CMS experiments at the LHC. This review has summarised the most recent results, based on the full run-2 dataset, corresponding to a luminosity of about 140~\ifb. A wide variety of fermionic final states has been probed: single-leptons, dileptons and multileptons; dijets  and multijets; as well as mixed leptonic and hadronic final states. Since no new-physics signals have been observed, ATLAS and CMS have extended their searches to BSM models with rarer production modes and more complex topologies, probing unexplored corners of parameter space. This has been made possible via the development of sophisticated multivariate tools, both in the reconstruction and identification of complex objects, such as tau leptons and heavy-flavour jets, and in the separation of potential signals from SM backgrounds. There has also been an increasing focus on generic model-independent searches, not tied to a particular BSM model.  The physics reach is summarised in tables~\ref{tab:lepton_channels}-\ref{tab:mixed_channels} and in figure~\ref{fig:summaryPlot} for a selection of benchmark scenarios. With this in mind, searches for fermionically-decaying heavy resonances have a promising future, with many exciting new results, and perhaps even discoveries,  expected in the near future from the ongoing LHC run-3 data (and its combination with the run-2 results) and in the next decades from the HL-LHC and possible future colliders.

\section*{DISCLOSURE STATEMENT}
The authors are not aware of any affiliations, memberships, funding, or financial holdings that might be perceived as affecting the objectivity of this review. 

\section*{ACKNOWLEDGMENTS}
The authors acknowledge the support of their respective national funding bodies, the UK STFC (CG) and Belgian F.R.S.-FNRS (BC). For the purpose of open access, the authors have applied a Creative Commons attribution (CC BY) license to any Author Accepted Manuscript version arising.  The authors thank ATLAS and CMS for providing their summary plot code. The summary plot in section~\ref{sec:summary_reach} was produced using a modification on the ATLAS code developed by St\'ephane Willocq.  

\bibliographystyle{ar-style5}
\bibliography{references}


\end{document}